\documentclass[journal=jctc,manuscript=article]{achemso}

\usepackage[utf8]{inputenc}
\usepackage[T1]{fontenc}
\usepackage{mathptmx}
\usepackage{wrapfig}                    
\usepackage{graphicx}				    
\usepackage{dcolumn}                    
\usepackage{amssymb, amstext, amsmath}  
\usepackage{physics}                    
\usepackage{bm}                         
\usepackage{nicefrac}                   
\usepackage{tabularx,booktabs}          
\usepackage{soul,xcolor}                
\usepackage{verbatim}                   
\usepackage{float,placeins}             
\usepackage{dsfont}
\usepackage{blkarray}
\usepackage{setspace}                   
\SectionNumbersOn

\title[\\]{Modeling surface vibrations and their role in molecular adsorption: a generalized Langevin approach}

\author{Ardavan Farahvash}
\affiliation{Department of Chemistry, Massachusetts Institute of Technology, Cambridge, Massachusetts 02139, USA}

\author{Mayank Agrawal}
\affiliation{School of Engineering, Brown University, Providence, Rhode Island 02912, USA}

\author{Andrew A. Peterson}
\affiliation{School of Engineering, Brown University, Providence, Rhode Island 02912, USA}

\author{Adam P. Willard}
\email{awillard@mit.edu}
\affiliation{Department of Chemistry, Massachusetts Institute of Technology, Cambridge, Massachusetts 02139, USA}

\setstcolor{red}

\begin{document}

\singlespacing

\begin{abstract}
    The atomic vibrations of a solid surface can play a significant role in the reactions of surface-bound molecules, as well as their adsorption and desorption.
    Relevant phonon modes can involve the collective motion of atoms over a wide array of length scales.
    In this manuscript, we demonstrate how the generalized Langevin equation can be utilized to describe these collective motions weighted by their coupling to individual sites. 
    Our approach builds upon the generalized Langevin oscillator (GLO) model originally developed by Tully.
    We extend the GLO by deriving parameters from atomistic simulation data. 
    We apply this approach to study the memory kernel of a model platinum surface and demonstrate that the memory kernel has a bimodal form due to coupling to both low-energy acoustic modes and high-energy modes near the Debye frequency. The same bimodal form was observed across a wide variety of solids of different elemental compositions, surface structures, and solvation states.
    By studying how these dominant modes depend on simulation size, we argue that the acoustic modes are frozen in the limit of macroscopic lattices. 
    By simulating periodically replicated slabs of various sizes we quantify the influence of phonon confinement effects in the memory kernel and their concomitant effect on simulated sticking coefficients.
\end{abstract}

\maketitle 
\section{Introduction}
    \label{sec:intro}
    Heterogeneous catalysis, and surface chemistry more broadly, is an intricate dance between the electronic and nuclear degrees of freedom of the molecular reagents and underlying substrate.
    An aspect of this dance that is sometimes overlooked is the vibrational modes of the substrate surface atoms.
    Indeed, several recent studies have highlighted that these modes can significantly influence reactions at metal interfaces \cite{tetenoire_why_2022,alducin_electrons_2019,shakouri_analysis_2018,von_boehn_promotion_2020}. 
    Methods that enable the study of these influences therefore have the potential to enhance our understanding of many surface chemical processes.
    One such method involves the generalized Langevin equation, which can reduce the complex dynamics of collective atomic surface motions to a single function: the memory kernel (or non-Markovian friction). 
    The memory kernel is a time-domain description of the phonon density of states re-weighted by the coupling of each phonon mode to surface atoms. 
    In this paper, we demonstrate how atomistic data and simulations can be used to parameterize the memory kernel and derive insights from its functional form. Crucially, we show that the memory kernel has a bimodal form arising from the strong coupling of surface sites to both coherent acoustic oscillations as well as a manifold of modes near the Debye frequency.
    This form is modified by the maximum phonon wavelength set by the simulation's boundary conditions in ways that can significantly affect surface chemistry, as we illustrate by calculating sticking coefficients. Specifically, nanoscale boundary conditions, such as those of atomistic simulation, cause the acoustic modes to oscillate on chemical timescales. We show that this effect can lead to a systematic decrease in sticking coefficients from the macroscopic limit. A GLE-based approach, such as we present here, can mitigate these unwanted finite-size effects without necessitating large and computationally expensive simulations.
    
    Over the past four decades, a multiplicity of experimental and theoretical studies have demonstrated how surface vibrations modulate reactions dynamics at metal interfaces. 
    Since at least the 1980's it has been understood that laser-induced photo-excitation of a metal surface can activate surface desorption \cite{stair_pulsed-laser-induced_1987}. The explanation of this effect has largely been attributed to hot electrons dissipating energy into the phonon-modes of the metal, which subsequently couple to molecular desorption \cite{stair_pulsed-laser-induced_1987,hall_pulsed-laser-induced_1987,bonn_phonon-_1999,frischkorn_femtochemistry_2006,gladh_electron-_2013,alducin_electrons_2019,tetenoire_why_2022}. 
    Another interesting case study has been the dissociation of methane on a variety of FCC metal surfaces. Several papers have demonstrated that the response of the lattice tunes the tunneling barrier for the dissociation of the carbon-hydrogen bond, leading to higher dissociation rates more consistent with experiment measurements than if one used a fixed surface \cite{luntz_activation_1989,luntz_ch4_1991,henkelman_theoretical_2001,nave_methane_2007,nave_methane_2009,nave_methane_2010,tiwari_methane_2009,tiwari_temperature_2010}. 
    Similar effects were observed for the dissociation of $\text{N}_{\text{2}}$ on Ru(0001), however in that case the reaction dynamics could still be well-described by treating the nuclei classically \cite{egeberg_molecular_2001,diekhoner_n2_2001,nattino_n2_2015,shakouri_accurate_2017,shakouri_analysis_2018}. 
    Separately, it has been noted that surface acoustic waves generated via piezoelectric drivers, can, when applied at specific frequencies and polarizations, strongly activate otherwise unfavorable reactions (such as the oxidation of ethanol on Pd) \cite{inoue_effects_2007,von_boehn_promotion_2020}.
    Inspired by both the wealth of work demonstrating the variety of ways surface vibrations influence chemical dynamics, we have revisited the GLE as a tool for understanding and simulating surface atom dynamics.

    The generalized Langevin equation describes the non-Markovian stochastic dynamics of a particle in a condensed phase environment as,
    \begin{equation}
        \label{eq:GLE}
        \dot{p}(t) = - \frac{d W}{d q}(t) - \int_{0}^{t} K(t-\tau) p(\tau) d\tau + R(t),
    \end{equation}
    where $q$ and $p$ represent the particle's position and momentum, respectively, and $W$, $K$, and $R$ are functions that characterize the influence of the  environment on particle dynamics.
    Specifically, $W$ is the potential of mean force, a function of particle position that encodes all relevant spatial correlations between the environment and the particle, $R$ represents the random forces on the particle, typically assumed to be time-correlated Gaussian noise, and $K$ is the memory kernel, which characterizes the time-correlated influences from the environment.
    $K$ is a time-dependent analog of the Markovian friction constant and is related to $R$ via the second fluctuation dissipation theorem (FDT) $ K(t) = \left \langle R(t) R(0)\right \rangle / (m k_\text{B} T) $. 
    As we elaborate upon throughout the rest of the paper, $K$ is of central importance to determining the dynamics of the GLE.
    
    Adelman and Doll\cite{adelman_generalized_1976} were the first to discuss in depth how the GLE could be used to model the effect of substrate phonons on a site within a purely harmonic lattice. Later, Tully developed the generalized Langevin oscillator (GLO) method, wherein the motion of a surface site is described via a GLE with a memory kernel that is given by a single exponentially damped sinusoid \cite{tully_dynamics_1980}. 
    As we discuss in Section \ref{subsec:extended_var}, such a memory kernel is equivalent to coupling the surface atom to a single dissipative (ghost) oscillator. 
    Tully's GLO method has seen much success as a highly computationally efficient way of modeling surface dynamics, particularly in application to molecular beam scattering experiments \cite{busnengo_trapping_2004,bukas_fingerprints_2015,nattino_modeling_2016,zhou_dynamics_2020,zhou_modified_2019}. 
    However, modeling the dynamics of the surface with a single mode is a limiting approximation; in principle surface atoms should couple to each normal mode of the lattice. 
    
    In this study, we extend Tully's GLO model to allow for a memory kernel of arbitrary complexity, and examine when and how the properties of the memory kernel affect molecular adsorption and surface scattering. We call this model the lattice generalized Langevin equation (LGLE). In order to parameterize the memory kernel we use data taken from atomistic simulations. Crucially, in Section \ref{sec:memory_results} we show that the qualitative properties of the memory kernel are independent of the atomistic model details, such as force field parameters.
    
    The remainder of the paper is organized as follows. In Section \ref{sec:theory}, we present the formal theory behind the LGLE, focusing on methods to parameterize the equation and briefly reviewing the extended variable transformation used to map the non-Markovian dynamics to a bath of dissipative harmonic oscillators. 
    In Section \ref{sec:methods}, we detail our simulation methods and in Section ~\ref{sec:memory_results}, we analyze memory kernels derived from these simulations, varying the choice of forcefields, elemental composition, solvation state, and lattice size. Finally, in Section \ref{sec:rate_results} we discuss how the properties of the memory kernel generated via our approach affect adsorption, highlighting systematic errors that occur in sticking coefficients when basing the memory kernel on data from small periodic systems. 

    \FloatBarrier

\section{Theoretical Background}
    \label{sec:theory}
    The LGLE treats the motion of surface sites using Eq.\ref{eq:GLE}, coarse graining over all other degrees of freedom in the solid/lattice. Interactions between surface sites and any molecular reagants are handled explicitly. The noise and memory terms in Eq.\ref{eq:GLE} arise from the influence of the bulk lattice on the surface atoms. 
    
    There are many approaches to parametrizing Eq.\ref{eq:GLE} for surface sites from atomistic information, and in this manuscript, we will discuss two in particular, henceforth termed the projection operator (PO) method and the correlation function (CF) method. The PO method is based on the approach detailed in the formative works of Adelman and Doll,\cite{adelman_generalized_1976} and Tully, \cite{tully_dynamics_1980} and involves matrix operations applied to the mass-weighted Hessian. 
    By contrast, the CF method requires the computation of time-correlation functions using MD (molecular dynamics) simulations, is commonly used in applications of the GLE to amorphous systems or liquid solutions \cite{canales_generalized_1998,gordon_generalized_2009,jung_iterative_2017,li_computing_2017,wang_implicit-solvent_2019}. Both approaches are useful and will be employed in subsequent sections.
    
    
    \subsection{Projection operator method}
    \label{subsec:po}

    The projection operator method is a simple and easily interpretable way of deriving and parameterizing the LGLE.
    The method requires computing and diagonalizing the mass-weighted Hessian of the substrate lattice.

    To begin, the many-body potential energy surface of the lattice $U(q_1 \ldots q_N)$ is expanded to second order about a local minimum,
    \begin{equation}
        U(q_1 \ldots q_N) \approx U_{0} + \frac{1}{2} \sum_{ij}^N \frac{\partial^2 U} {\partial q_i \partial q_j} q_i q_j
    \end{equation}
    where $U_{0}$ is the potential energy at the minimum-energy configuration, $\{q_i\}$ is the set of displacements of each atom from its equilibrium position, and $N$ is the total number of atoms in the system.
    In this expansion, the force along any displacement is $F_k = -(\partial U/\partial q_k) = -\sum_i \frac{\partial^2 U}{\partial q_i \partial q_k} q_i$.
    By introducing mass-weighted coordinates $x_i = \sqrt{m_i}q_i$ and the mass-weighted Hessian $D^2_{ij} = \frac{1}{\sqrt{m_im_j}} \frac{\partial^2 U}{\partial q_i \partial q_j}$ the resulting equation of motion may be written as,  
    \begin{equation}
        \label{eq:harmonic_eom}
        \ddot{ \mathbf{x} } = - \mathbf{D}^2 \mathbf{x}.
    \end{equation}
    
    Two projection operators, $\mathbf{P}$ and $\mathbf{Q}=\mathds{1} - \mathbf{P}$ are then used to separate this equation in to system (surface site) and bath (remaining lattice) subspaces respectively,
    \begin{equation}
        \label{eq:proj_p}
        \ddot{ \mathbf{x} }_P = -\mathbf{D}^2_{PP} \mathbf{x}_P -\mathbf{D}^2_{PQ} \mathbf{x}_Q,
    \end{equation}
    \begin{equation}
        \label{eq:proj_q}
        \ddot{ \mathbf{x} }_Q  = -\mathbf{D}^2_{QP} \mathbf{x}_P -\mathbf{D}^2_{QQ} \mathbf{x}_Q,
    \end{equation}
    where $\mathbf{x}_P= \mathbf{Px}$ are the system degrees of freedom, $\mathbf{x}_Q= \mathbf{Qx}$ are the bath degrees of freedom and $\mathbf{D}^2_{PP} = \mathbf{P D^2 P}$, etc.  In principle, these projection operators can take any form, so long as they obey the properties of idempotency $\mathbf{P}^2 = \mathbf{P}$ and orthogonality $\mathbf{PQ}=0$. However, if we wish $\mathbf{x}_P$ to correspond to the displacement of a particular surface atom, it is most natural to choose $\mathbf{P}$ to be a matrix with ones on the diagonal for the indices corresponding to the coordinate(s) of interest and zeros elsewhere,
    \begin{equation}
        \mathbf{P}=
        \left(
        \begin{array}{c|c}
        \mathds{1} & 0 \\ \hline
        0 & 
        \begin{array}{c c c} 
              &   &  \\ 
              & 0 &  \\ 
              &   &  
        \end{array}
        \end{array}
        \right)
        \qquad 
        \mathbf{D}^2=
        \left(
        \begin{array}{c|c}
        \mathbf{D}^2_{PP} & \mathbf{D}^2_{PQ} \\ \hline
        \mathbf{D}^2_{QP} & 
            \begin{array}{c c c} 
                  &   &  \\ 
                  & \mathbf{D}^2_{QQ} &  \\ 
                  &   &  
          \end{array}
        \end{array}
        \right).
    \end{equation}
    
    For simplicity, let us allow $\mathbf{x}_P$ to be a scalar $x_P$ corresponding to the displacement of a single surface atom along a single coordinate. Concomitantly, if we let $N$ be the total number of sites in the lattice, $\mathbf{D}^2_{PQ}$ becomes an $1 \times (N-1)$ row vector,  $\mathbf{D}^2_{QP}$ becomes an $(N-1) \times 1$ column vector, and $\mathbf{D}^2_{PP}$ becomes a $1 \times 1$ scalar and thus is henceforth unbolded. 
    
    We then proceed by solving Eq.~\ref{eq:proj_q} in terms of $x_P$, substituting that solution into Eq.~\ref{eq:proj_p}, and subsequently transforming back from mass-weighted to standard coordinates. 
    The final solution is of the form of Eq.~\ref{eq:GLE} and describes the motion of a surface atom with,
    
    \begin{equation}
        \label{eq:po_K}
        K(t) = \mathbf{D}^2_{PQ} \mathbf{D}^{-2}_{QQ} \cos( \mathbf{D}_{QQ} t) \mathbf{D}^2_{QP},
    \end{equation}
    \begin{equation}
        \label{eq:po_W}
        W(q) = m\left[-D^2_{PP} + K(t{=}0)\right]q^2,
    \end{equation}
    \begin{equation}
        \label{eq:po_R}
        \frac{R(t)}{\sqrt{m}} = - \mathbf{D}^2_{PQ} 
        \left[  
           \cos(\mathbf{D}_{QQ} t) \mathbf{x}_Q(0)  
        +  \mathbf{D}^{-1}_{QQ}\sin(\mathbf{D}_{QQ} t) \mathbf{\dot{x}}_Q(0) 
        +  \mathbf{D}^{-2}_{QQ}\cos(\mathbf{D}_{QQ} t) \mathbf{D}^2_{QP} \mathbf{x}_P(0) 
        \right].
    \end{equation}
    Disregarding some mathematical subtleties with respect to the third term in Eq.~\ref{eq:po_R}\cite{hanggi_generalized_1997}, $K(t)$ and $R(t)$ above do satisfy the FDT. Note that because $K(t)$ determines the properties of $R(t)$ via the FDT and also the determines the deviation of $W$ from a fixed lattice, $K(t)$ is arguably the most fundamental quantity in Eq.~\ref{eq:GLE}, containing all the relevant information for how the bath modulates the system's dynamics.
    Throughout the rest of this paper, it will prove useful to analyze the Fourier transform of the memory kernel, $\bar{K}(\omega)$, which can be shown to be equivalent to the power spectrum of the noise $R(t)$ by the Wiener-Khinchine theorem,
    \begin{equation}
        \label{eq:power_spectrum}
        \bar{K}(\omega) = \sum_{i} \frac{c^2_i}{\omega^2_i} \delta(\omega - \omega_i).
    \end{equation}
    In the equation above, $\omega_i$ are the normal mode frequencies of the bulk lattice Hessian $\mathbf{D}^2_{QQ}$, and $c_i$ are the coupling constants between the surface degree of freedom and $i$th normal mode, $c_i = \sum_{j} [\mathbf{D}^2_{PQ}]_{1j} V_{ji} $, where $\mathbf{V}$ is a matrix with the eigenvectors of $\mathbf{D}^2_{QQ}$ as columns. The index $j$ runs over the rows of $V$ and $i$ over columns.
    Note that we ignore the negative frequency components of the Fourier transform in Eq.~\ref{eq:power_spectrum} as they are simply a reflection of the positive components. Eq.~\ref{eq:power_spectrum} reveals that the peaks of $\bar{K}(\omega)$ are nothing more than the phonon frequencies of the lattice weighted by their relative coupling to the site of interest. We will use this powerful interpretation throughout the rest of our paper. 
    
    \subsection{Correlation function method}
    \label{subsec:cf}
    A notable disadvantage of the projection operator method outlined in the previous section is that it relies on the assumption that the PES is harmonic (Eq.~\ref{eq:harmonic_eom}).
    Unfortunately, some systems exhibit significant anharmonicities, such as solvated surfaces or surfaces with defects.
    In cases such as these, the GLE parameters can be determined by analyzing system correlation functions. By system we are referring to the degrees of freedom of interest, which for our purposes correspond to the coordinates of the surface sites of the lattice.
    This method begins from recognizing that the random force $R(t)$ must be uncorrelated with the system's momenta: $\left \langle R(t) p(0) \right \rangle = 0$.
    This identity can be considered prerequisite for $R(t)$ to be properly interpreted as a "random" noise, and is indeed consistent with Eq.~\ref{eq:po_R}. 
    
    By taking the equilibrium time correlation function of both sides of the generalized Langevin equation (Eq.~\ref{eq:GLE}) with the initial momentum we find,
    \begin{equation}
        \label{eq:volterra}
        \left \langle \dot{p}(t) p(0) \right \rangle 
        + \left \langle  \frac{d W}{d q}(t) p(0) \right \rangle
        = - \int_{0}^{t} K(t-\tau) \left \langle p(\tau) p(0) \right \rangle    d\tau.
    \end{equation}
    Using MD simulation the force-momentum correlation functions (left-hand side) and the momentum autocorrelation function may be computed, and subsequently Eq.~\ref{eq:volterra} may be solved to find $K(t)$. 
    Standard numerical methods, such as those based on the use of the fast Fourier transform algorithm and convolution theorems lack the numerical accuracy to reliably solve Eq.~\ref{eq:volterra} \cite{berne_calculation_1970,wang_implicit-solvent_2019}.
    The method we use to solve Eq.~\ref{eq:volterra} is described in detail in the supplementary information.

    \textcolor{black}{It should be noted that for a general, non-linear and/or time-dependent environment, the memory kernel is a non-linear operator and the noise term may not be Gaussian \cite{ayaz_generalized_2022}. In this paper we assume that a linear GLE with Gaussian noise to be sufficient model for the surface fluctuations of a solid lattice.}
    
    \subsection{Extended Variable Transformation}
    \label{subsec:extended_var}
    Computing the integral over the memory kernel in Eq.~\ref{eq:GLE} is computationally intensive, especially for systems with long correlation times.
    A common strategy for simplifying this calculation is to expand the GLE into a set of Markovian equations.
    These equations describe system bilinearly coupled to bath of dissipative, stochastic harmonic oscillators. 
    The advantage of using a dissipative bath, as opposed to an energy conserving bath, is that an extended bath can often be represented with one or two dissipative oscillators and that doing so dramatically reduces the dimensionality of the equations of motion.
    Here we briefly summarize the method, noting that excellent presentations of the same formalism can be found in Ref. \citenum{ceriotti_colored-noise_2010} and Ref. \citenum{baczewski_numerical_2013}.
    
    Given a GLE with a memory kernel that is a finite sum of exponentially damped sinusoids,
    \begin{equation}
        \label{eq:series_exp_gle}
        K(t) = \sum_{i=1}^N e^{-\gamma_i t} \left( C_i \cos(\omega_i t) + D_i \sin(\omega_i t) \right), 
    \end{equation}
    the original non-Markovian equation of motion can be replaced with,
    \begin{equation}
        \label{eq:ext_var_eom}
        \frac{d}{dt}
        \begin{pmatrix}
            p \\
            \mathbf{b} \\
        \end{pmatrix}
        = 
        \begin{pmatrix}
            - \frac{dW}{dq} \\
            0 \\
        \end{pmatrix}
        +
        \begin{pmatrix}
            0 & \mathbf{A}_{pb} \\
            \mathbf{A}_{bp} & \mathbf{A}_{b} \\
        \end{pmatrix}
        \begin{pmatrix}
            p \\
            \mathbf{b} \\
        \end{pmatrix}
        +
        \begin{pmatrix}
            0 & 0 \\
            0 & \mathbf{B}_{b} \\
        \end{pmatrix}
        \begin{pmatrix}
             \\
            d\mathbf{W} \\
        \end{pmatrix}.
    \end{equation}
    Here $p$ is the system's momenta, and $\mathbf{b}$ is a set of bath variables we must involve in time with our system. $d\mathbf{W}$ is an array of uncorrelated Gaussian random variables satisfying $\left \langle dW_i(t) dW_j(0) \right \rangle = \delta_{ij} \delta(t)$, where $\delta_{ij}$ is the Kronecker delta and $\delta(t)$ the Dirac delta. The matrix $\mathbf{A}_{b}$ is block diagonal with entries,
    \begin{equation}
        \label{eq:def_Abb}
        \mathbf{A}_{b} = 
        \begin{pmatrix}
            2 \gamma_i & \sqrt{\gamma_i^2 + \omega_i^2} \\
            -\sqrt{\gamma_i^2 + \omega_i^2} & 0 \\
        \end{pmatrix},
    \end{equation}
    and $\mathbf{A}_{pb}$  and $\mathbf{A}_{bp}$ are arrays of form,
    \begin{equation}
        \label{eq:def_Apb_Abp}
        \mathbf{A}_{pb} = 
        \begin{pmatrix}
            \sqrt{\frac{C_i}{2} - 2 \frac{ D_i \omega_i^2}{\gamma_i} } & 
            \sqrt{\frac{C_i}{2} + 2 \frac{ D_i \omega_i^2}{\gamma_i} }
        \end{pmatrix},
        \qquad
        \mathbf{A}_{bp} = 
        \begin{pmatrix}
            \sqrt{\frac{C_i}{2} - 2 \frac{ D_i \omega_i^2}{\gamma_i} } \\ 
            \sqrt{\frac{C_i}{2} + 2 \frac{ D_i \omega_i^2}{\gamma_i} }
        \end{pmatrix}.
    \end{equation}
    The matrix $\mathbf{B}_{b}$ is related to $\mathbf{A}_{b}$ by the equation, 
    \begin{equation}
        \label{eq:extvar_FDT}
        \mathbf{B}_{b} \mathbf{B}^T_{b} = k_\mathrm{B} T (\mathbf{A}_{b} + \mathbf{A}_{b}^T),
    \end{equation}
    which ensures that the ensuing dynamics obey the fluctuation-dissipation theorem. 
    
    Tully's GLO model is based on the same approach beginning from Eq.~\ref{eq:series_exp_gle} and setting the number of terms $N=1$.
    In the work we present in this manuscript, we determine the optimal values of $N$, as determined based on the analysis of $K(t)$.
    We drop the sin terms in Eq.~\ref{eq:series_exp_gle}, thus casting the memory kernel as a set of exponentially damped cosines and yielding a Lorentzian power spectrum of the form,
    \begin{equation}
        \label{eq:lorentzian_spectral_density}
        \bar{K}(\omega) = \sum_{i=1}^N C_i \left( \frac{\gamma_i}{\gamma_i^2 + (\omega - \omega_i)^2} \right).
    \end{equation}

\section{Simulation Details}
    \label{sec:methods}
    Simulations using Effective Medium Theory (EMT)~\cite{Norskov1980} or Embedded Atom Method (EAM)~\cite{Daw1983} forcefields for metal dynamics were performed using the Atomic Simulation Environment \cite{larsen_atomic_2017,tadmor:elliott:2011,OpenKIM-MO:108408461881:001,OpenKIM-MD:128315414717:004,OpenKIM-MO:757342646688:000,OpenKIM-MD:120291908751:005}.
    The parameters for these forcefields were taken from Ref. \citenum{jacobsen_semi-empirical_1996} and Ref. \citenum{zhou_misfit-energy-increasing_2004} respectively. Simulations using Lennard-Jones forcefield, both solvated and in vacuum state, were performed using LAMMPS. Lennard-Jones forcefield parameters were taken from Ref. \citenum{heinz_accurate_2008}. The solvent used was SPC/E \cite{berendsen_missing_1987}. 
    
    All simulations were performed in two steps. First, a temperature equilibriation step was run for 50 picoseconds at 300K using a Langevin thermostat. Afterwards simulations were run in an constant energy ensemble using the velocity Verlet algorithm for 4 nanoseconds. Only data from the NVE step was used in subsequent analysis and calculations. All simulations were performed using periodic boundary conditions in the X/Y directions (parallel to the surface). Four atoms of the bottom row of the unit cell were rigidly constrained in order to remove center of mass motion. 
    
    For the surface scattering simulations used to generate data for Section \ref{sec:rate_results}, 5000 independent trajectories were averaged per value of the incident velocity to obtain sticking coefficients for GLE simulations, while 2000 independent trajectories were averaged for EMT simulations. Each trajectory was twenty picoseconds in length, which we found to be adequate for the convergence of results. 

\section{Memory kernels and power spectra for surface sites in metal lattices}
    \label{sec:memory_results}
    
    We begin by analyzing the memory kernel for the fluctuations of a single atom site in the surface of a 4x4x4 cell of Pt(111) with periodic boundary conditions. Results were calculated for each surface site individually and subsequently averaged. 
    All memory kernels and power spectra presented in the main text are calculated via the correlation function (CF) approach (Sec.~\ref{subsec:cf}). 
    We present results using the projection operator (PO) approach (Sec.~\ref{subsec:po}) in the supplementary information, and will refer these results when necessary in the main text. Evidence for the convergence of the memory kernels presented here is given in Fig.~S1 and Fig.~S2. 
    
    \begin{figure*}[h]  
       \centering
       \includegraphics*[width=5.25in]{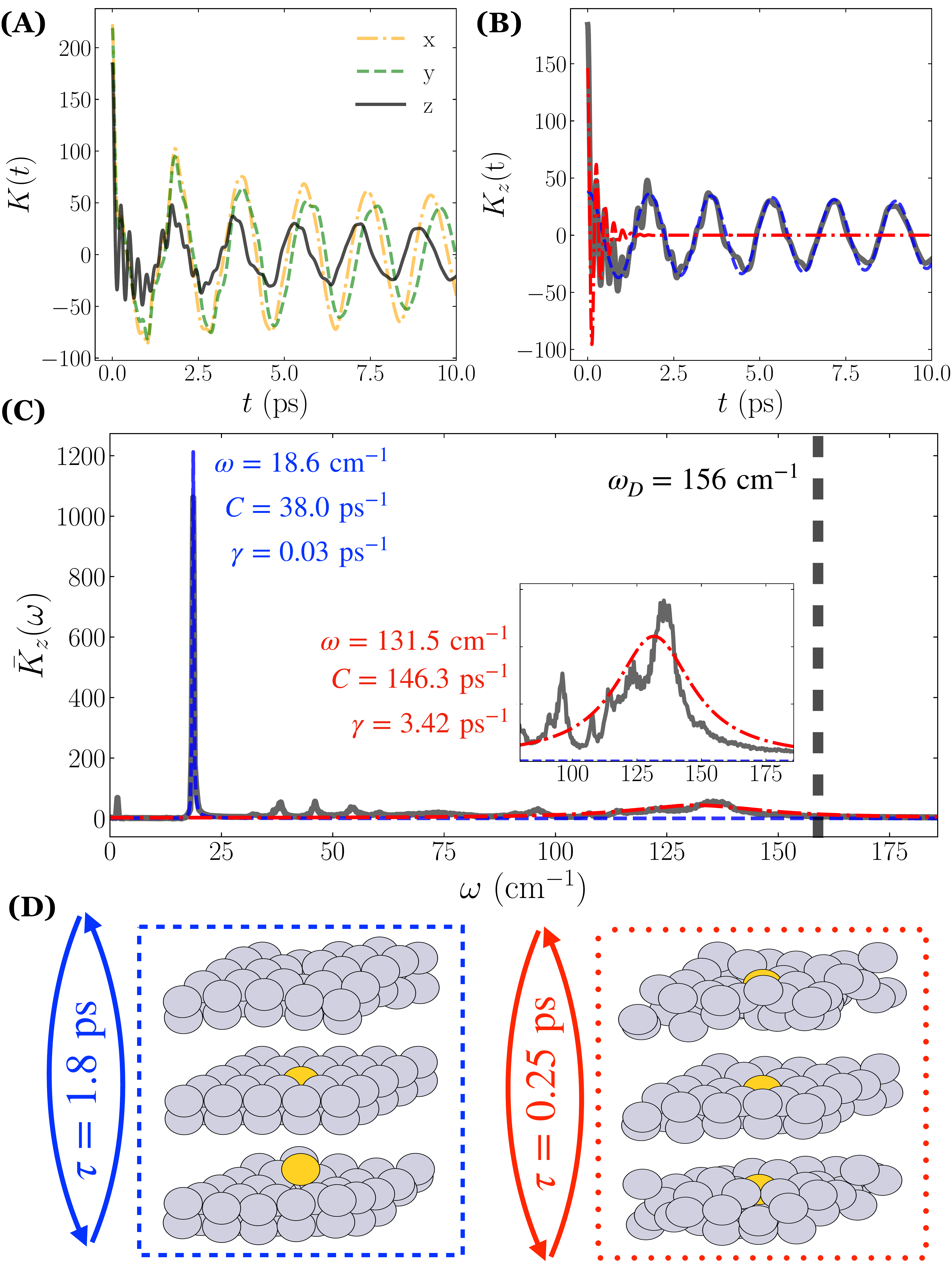}
        \caption{Memory kernel and random force power spectrum for surface sites of a Pt(111) lattice computed using an EMT forcefield. (A) Memory kernel for fluctuations in $x$/$y$ (in-surface plane) and $z$ (out of plane) directions. (B) $z$ component of the memory kernel. Red and blue lines are two exponential sinusoids optimized to fit the computed memory kernel (grey line). (C) Power spectrum of $z$ component of the memory kernel. The grey dashed vertical line corresponds to the experimental Debye frequency. (D) Depiction of lattice normal modes most associated with red and blue lines. $\tau$ is the period of the respective normal mode.
        \textcolor{black}{The yellow-highlighted sphere in the middle of the lattice represents the chosen adsorption site, which is fixed in space, while the surrounding grey atoms represent the fluctuating bulk lattice.}}

       \label{figure1}
    \end{figure*}
    Figure \ref{figure1}A presents the orthogonal $x$, $y$, and $z$ components of the memory kernel calculated from all-atom simulations using an Effective Medium Theory (EMT) forcefield. The $x$ and $y$ components arise from fluctuations in the plane of the lattice and the $z$ component arises from fluctuations normal to it. Note the anisotropy between the $x$ and $y$ components and the $z$ component, a simple consequence of the difference in the number of the coordination number and anisotropy between longitudinal and transverse modes. 
    In the analysis that follows, we focus only on $K_z$---the component of the memory kernel acting in the $z$ coordinate---because fluctuations in this direction (perpendicular to the metal surface) are most relevant to surface adsorption/desorption.
    
    In Figure \ref{figure1}C we present the noise power spectrum (Fourier transform of the memory kernel) of the $z$ component $K_z(\omega)$ specifically. 
    As elaborated upon in Section \ref{sec:theory}, each peak in the power spectrum gives information about the lattice phonon modes and how they couple to the motion of a surface site. 
    The value of $\omega$ at each peak corresponds to the frequency of the mode, the width $\gamma$ corresponds to the timescale of energy exchange or dissipation between the mode and the surface site, and the coefficient $C$ indicates the coupling strength of the system to that particular mode.

    The power spectrum in Figure \ref{figure1}C is essentially bimodal. 
    The red peak---henceforth referred to as the acoustic peak---is centered at a low frequency ($\omega = 18.6 \text{cm}^{-1}$) and thus exchanges energy relatively slowly ($\gamma = 0.03 \text{ps}^{-1}$), while the blue peak---henceforth referred to as the Debye peak---is centered near the Debye frequency of Pt ($\omega = 131 \text{cm}^{-1}$) and exchanges energy relatively rapidly ($\gamma = 3.42 \text{ps}^{-1}$). 
    By comparing the power spectrum computed via the CF method to the power spectrum computed via the PO method (Figure~S4), it is possible to determine precisely which normal modes of the lattice are primarily responsible for these two peaks. 
    These normal modes are illustrated in Figure \ref{figure1}D. The acoustic peak arises from a longitudinal acoustic oscillations normal to the surface plane. Meanwhile, the Debye peak arises from many closely spaced normal modes near the Debye frequency, which consist of atomic scale local oscillations.

    \begin{figure*}[h]  
       \centering
       \includegraphics*[width=3.25in]{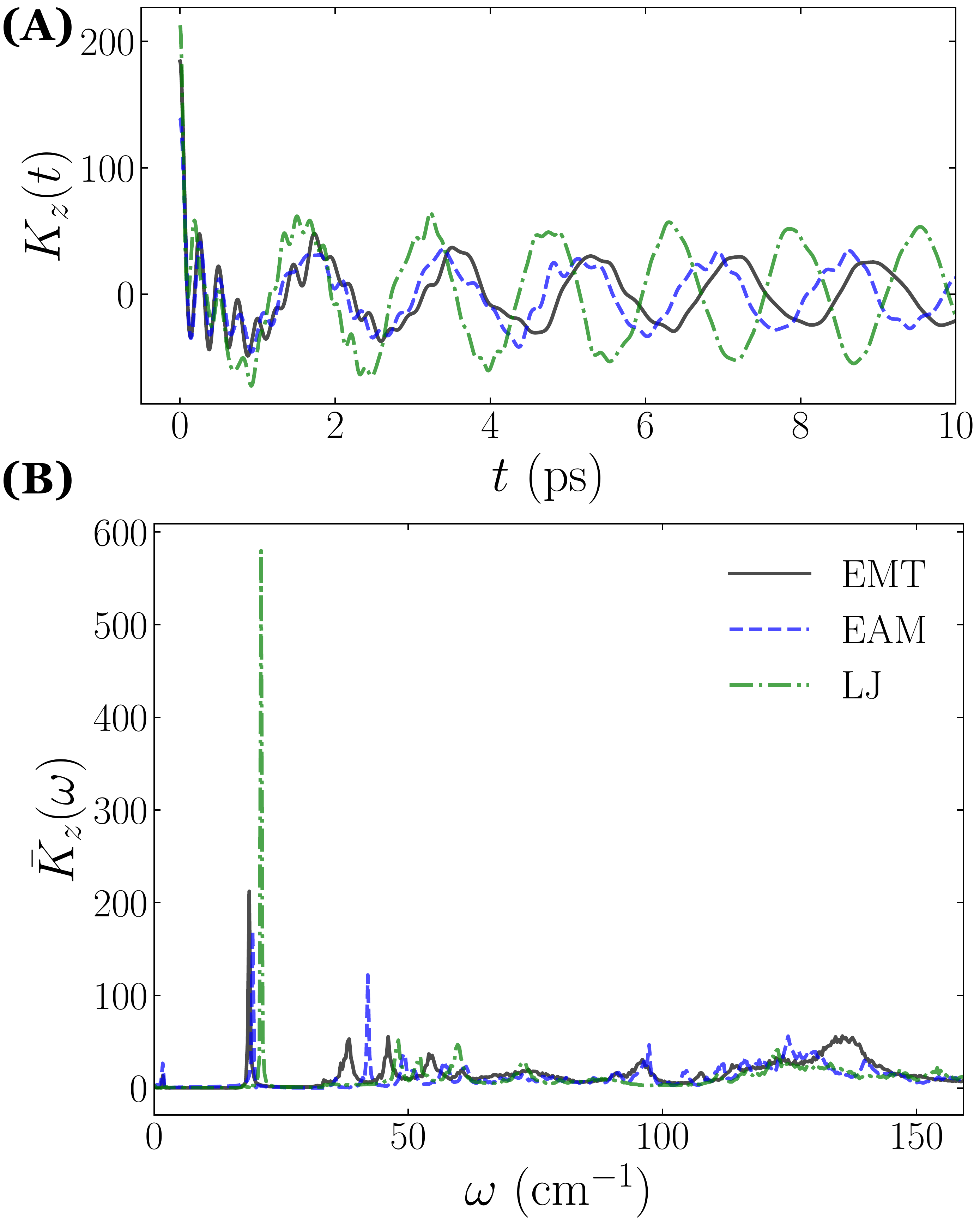}
       \caption{ (A) Memory kernel and (B) power spectrum for surface site fluctuations of Pt(111) simulated using three different atomistic models: Effective Medium Theory, Embedded-Atom Method, and a Lennard-Jones model. The acoustic peak at 20 cm$^{-1}$ of the power spectrum has been scaled by a factor of 0.3 for the purposes of visual clarity.}
       \label{figure2}
    \end{figure*}
    In order to ensure the validity and transferability of our results, we tested the forcefields other than EMT. These results are illustrated in  (Figure \ref{figure2}). The Lennard-Jones (LJ) model is based on a very different underlying physics than the EMT/EAM models (LJ model uses only pairwise interactions, while EMT/EAM are both many-body potentials based on the local atom density). Despite this fact, all three models produce the same qualitative bimodal form. 
    Much of the quantitative difference between the EMT/EAM and LJ models can be explained by the fact that the LJ model results in a stiffer lattice than  EMT and EAM. 
    The lattice stiffness can be roughly quantified in terms of the average value of the mass-weighted Hessian $k_{lat} = \langle \mathbf{D}^2 \rangle$. For EMT the stiffness of the 4x4x4 Pt(111) lattice is 19.5 kJ/(mol nm\textsuperscript{2}), for EAM it is 19.8 kJ/(mol nm\textsuperscript{2}), and for LJ it is 37.1 kJ/(mol nm\textsuperscript{2}).

    We also tested lattices of different elemental composition and surface facets (Figure~S6). Once again, although variations were observed in the location, widths, and heights of the primary peaks of the power spectrum, all of the lattices exhibited the same qualitative bimodal response. 
    The apparent universality of this bimodal behavior might be considered a simple consequence of the structure of Eq.~\ref{eq:power_spectrum}. Essentially, there is a trade-off between the amplitude of each mode, which decreases as the inverse square of the frequency, and the density of modes, which increases sharply near the Debye frequency.

    However, it is worth noting that this universality is not trivial. The bimodal behavior is not recovered in simple 1D systems with nearest-neighbor interaction (see Ref. \citenum{florencio_exact_1985} and \citenum{lee_local_2016} and Section S4 of the Supplementary Material for more details), and therefore is an emergent property of the 3D metal lattice. 

    \subsection{Finite-size effects}
\begin{figure*}[h]  
       \centering
       \includegraphics*[width=5.25in]{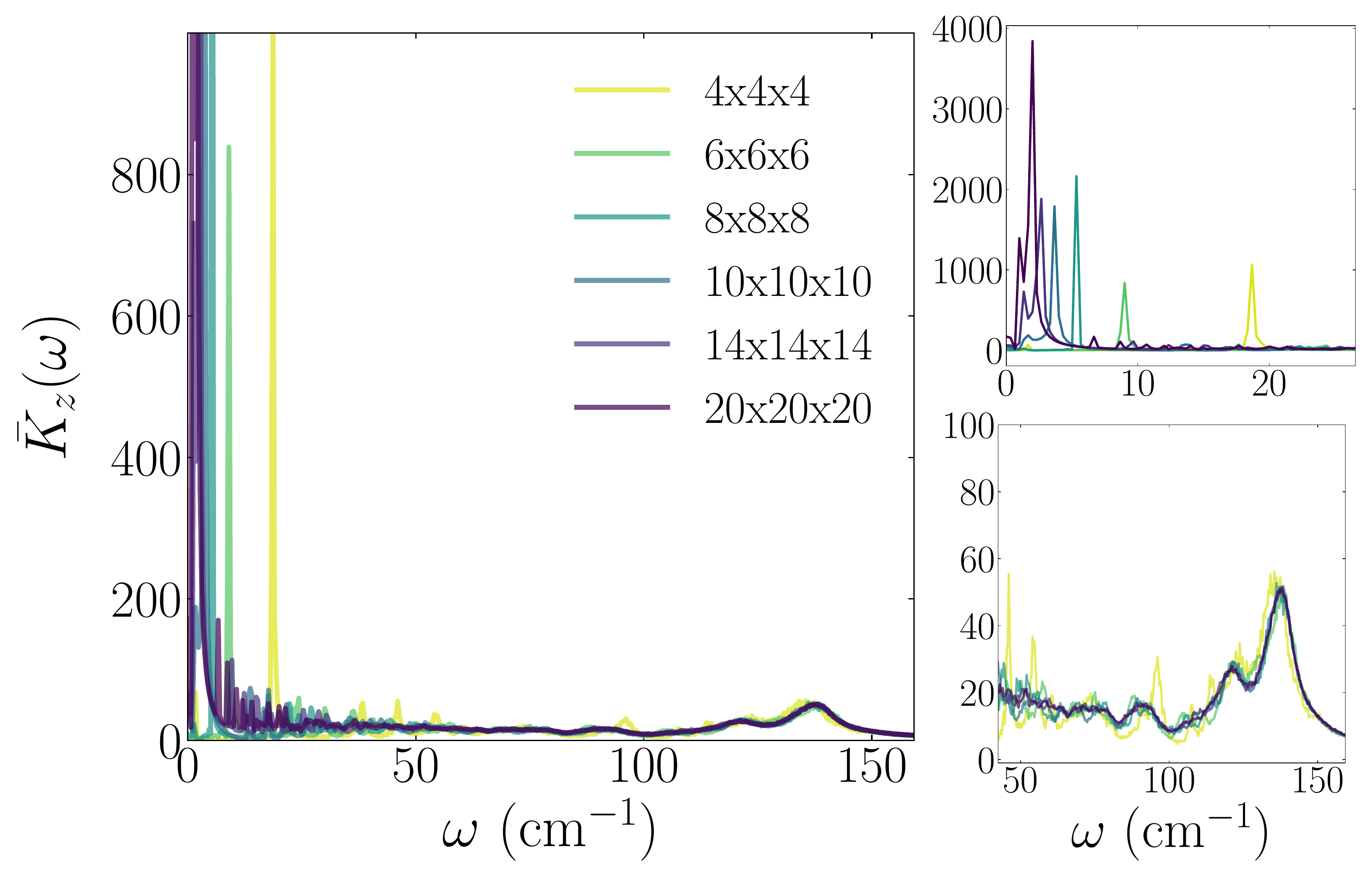}
       \caption{Memory kernel and power spectra for surface site fluctuations of Pt(111) lattices of different sizes. The plots on the right zoom-in on the regions near to acoustic and Debye peaks respectively}
       \label{figure3}
    \end{figure*}

    Periodic boundary conditions are used in MD simulations to mimic an infinite system but can still introduce finite size effects into phonon-mediated processes.
    Specifically, the size of the crystalline solid sets the maximum phonon wavelength. 
    As such, increasing the size of the simulation should shift the frequencies of the acoustic modes, which in turn affects the shape of the memory kernel.
    In Figure \ref{figure3} we confirm that the acoustic peak shifts to lower frequency as the size of the lattice increases, while the Debye peak remains unchanged.
    Even when increasing the lattice size of $\sim$45 nm (8000 atoms), the memory kernel and power spectrum do not converge.
    In fact, the frequency ratio between the acoustic peak of different size lattices roughly agrees with the results of an isotropic wave equation, suggesting that the acoustic peak will never converge to a fixed frequency, but rather decrease like as $1/L$ where $L$ is the side-length of the lattice.

    The demonstration that the memory kernel exhibits significant finite size effects has implications for the accuracy of simulation studies of surface phenomena.
    Perhaps most notably, this demonstration suggests than in the macroscopic limit, the frequencies of acoustic mode will be too low to affect any chemical dynamics at the surface. 
    In other words, the acoustic mode are effectively frozen.
    Therefore, for observables that depend on memory all finite-size simulations contain an intrinsic error which is purely kinetic in nature. 
    We demonstrate this explicitly in Section \ref{sec:rate_results}.
    
    The size-dependence of the memory kernel may also have ramifications for nanoparticle catalyst design, because it demonstrates how nanoparticle vibrational modes behave quite differently than their macroscopic counter parts, much as their electronic modes do. Indeed, experimental studies of electron relaxation in metal supported nanoparticles have already shown that the phonon-mediated dissipation of electron energy depends strongly on the nanoparticle size\cite{arbouet_electron-phonon_2003}.

    \subsection{Solvation Effects}
    \begin{figure*}[h]  
       \centering
       \includegraphics*[width=3.25in]{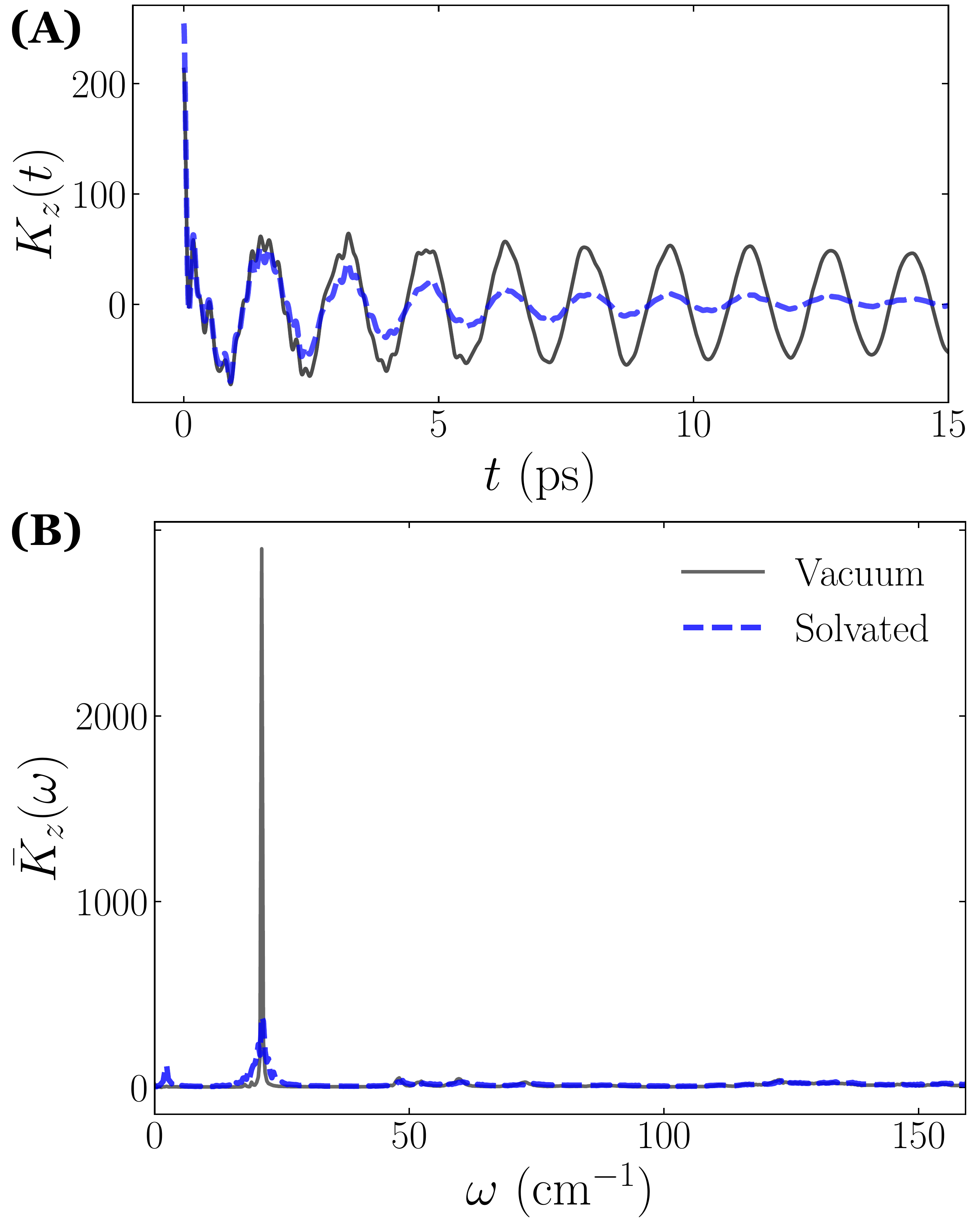}
       \caption{(A) Memory kernel and (B) noise power spectra for surface site fluctuations of Pt(111) simulated using a LJ model with and without SPC/E solvent.}
       \label{figure4}
    \end{figure*}
    We now consider the effects of an adjacent solvent on the memory kernel of surface lattice vibrations.
    Thus far, we have assumed an idealized lattice in the limit of low pressure and substrate surface coverage. 
    Most catalysts operate under conditions where there is significant surface coverage by solvent and/or reactive species. 
    Here we will explore how solvation affects surface site fluctuations by computing memory kernels for Pt(111) surfaces solvated in SPC/E water using the CF approach. We save the more difficult, yet still very important question, of surface coverage by physisorbed and chemisorbed species for future study.
    
    Figure~\ref{figure4} demonstrates the difference between Pt(111) surfaces in vacuum versus in solvent. The primary difference is in the damping of the acoustic mode, whose coupling to the surface site motion is much smaller when the surface is solvated. This effect is likely attributable to the additional pressure exerted by the solvent, making large fluctuations in the direction normal to the surface plane more energetically costly. The damping of the acoustic mode suggests that the finite-size effects discussed previously are likely far less important for solvated surfaces than they are for surfaces in gas-phase.  

\section{How memory affects adsorption and scattering}
    \label{sec:rate_results}
    
    Molecular beam scattering experiments are an invaluable tool for understanding the properties of surface reactions, elucidating information about binding potential energy surface and energy dissipation rates of the lattice \cite{sitz_gas_2002,juurlink_state-resolved_2009,chadwick_quantum_2017}. When the incident particles are sampled from an appropriate thermal distribution, the surface sticking probability can be shown to be proportional to the adsorption rate constant. Tully's GLO model is often used in simulations of surface scattering (either reactive or non-reactive) as an efficient computational method for describing energy loss to the lattice during the scattering process \cite{busnengo_trapping_2004,bukas_fingerprints_2015,nattino_modeling_2016,zhou_dynamics_2020,zhou_modified_2019}. 
    In this section, we employ the LGLE for the same purpose, specifically studying the differences between the finite-size limit and the macroscopic limit (when the low-energy acoustic modes are held fixed). 
    
    We demonstrate our approach on the simple case of the non-reactive scattering of Argon on Pt(111), in the direction normal to the surface. 
    The PES describing the interaction between the argon atom and the Pt surface  is taken to be of Morse form,
    \begin{equation}
        \label{eq:morse_potential}
        U(\Delta z) = D\left [ 1-e^{-a(\Delta z-z_0)} \right ]^2,
    \end{equation}
    where $\Delta z$ is the vertical distance between the Ar and the nearest Pt atom, $D$ is a parameter which controls the depth of the PES well, $a$ controls the width of the well, and $z_0$ is the location of the well's minimum.
    The values of these parameters were fit from DFT calculations presented in Ref. \citenum{chen_role_2012} using a van der Waals density functional (vdW-DF2) and are presented in Table S1. 
    Each scattering trajectory was initialized outside the Morse well at a distance of $z=15$ Anstroms, with a fixed initial velocity directed into the well and randomized lattice atom positions/velocities \textcolor{black}{drawn from a thermal distirbution at 300K}.
    From these trajectories, the sticking probability, $S$, was calculated, as the ratio of trajectories that remain trapped within the well ($z \leq 15$) Angstroms after a collision with the surface. 

    \begin{figure*}[h]  
       \centering
       \includegraphics*[width=6.5in]{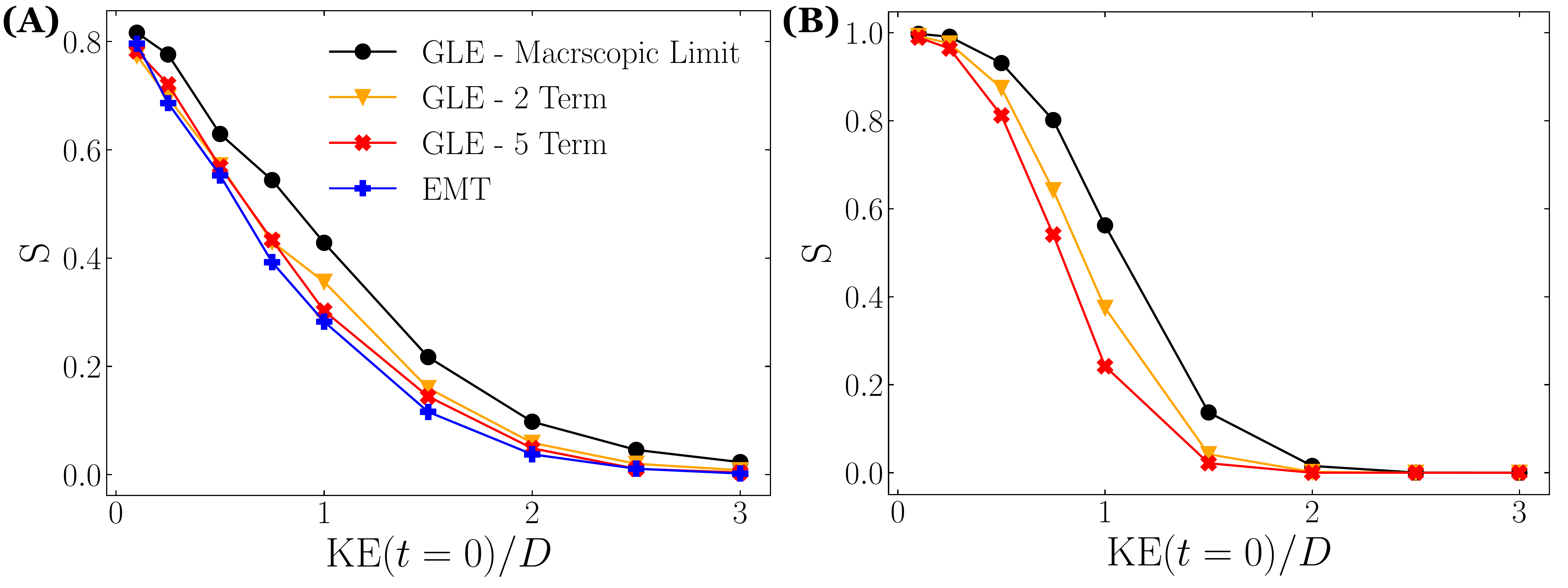}
       \caption{ Sticking probabilities $\mathrm{S}$ as a function of the ratio of the incident kinetic energy to the well depth $\mathrm{KE}(t=0)/D$. (A) Results for Morse PES with $D=6.62 \mathrm{eV}$. (B) Results for Morse PES with an increased well-depth, $D=30.62 \mathrm{eV}$. The blue curve uses an all-atom simulation using an EMT forcefield to treat the metal degrees of freedom. 
        The red and orange curves use the LGLE (Eq.~\ref{eq:GLE}) parameterized from a 4x4x4 EMT simulation to treat the metal. 
        The red curve uses only two damped sinusoids to fit $K(t)$, while the orange curve uses a five term fit give a more accurate estimation of the memory kernel and power spectrum (see Figure S7.). The black curve corresponds to the extrapolated macroscopic limit of the LGLE, wherein the surface site motion is coupled to only to the Debye mode.
        }
       \label{figure5}
    \end{figure*}
    
    Figure \ref{figure5}A illustrates variations the sticking probability using four different simulation approaches. 
    The blue, red, and orange curves of Figure \ref{figure5}A largely agree with one another, illustrating that the LGLE accurately captures the dynamics of the forcefield it is parameterized from. More interesting however, is the consistent increase in the sticking probability between the nanoscale lattices (either modeled with EMT or GLE) and the macroscopic limit. This discrepancy can be qualitatively explained by the relative dissipation rates of the acoustic and Debye modes. 
    Since nanoscale lattices couple the motion of surface atoms to the acoustic modes, and these acoustic modes dissipate energy much slower than the Debye mode, collisions with nanoscale lattices are more elastic. This effect can be observed more explicitly by studying histograms of the energy dissipated over many scattering trajectories (Figure S8). 
    
    In Figure \ref{figure5}B we study the scattering probability of a stickier particle with a well-depth that is nearly 5 times greater. Increasing $D$ increases the effective coupling between the adsorbate and the metal phonon bath, exacerbating the finite-size effects seen in Figure \ref{figure5}B. 
    The results in Fig.~\ref{figure5} highlights an error that is inherent to nanoscale simulations of surface-molecule interactions.
    This error is due purely to the phonon confinement effects placed by the boundary conditions, and can be exacerbated by errors in the adsorbate's binding energy to the metal. 
    
    Finally, we also tested how the simulation approaches examined in this section affect desorption rate constants using the same Morse potential (Eq.~\ref{eq:morse_potential}). We did so because it is well-known that the rate constants do depend sensitively on the character of the memory kernel. The results are presented in the SI (Figures S11 and S12). Interestingly, no significant difference was found in the desorption rate constant between LGLE models parameterized from nanoscale simulation and the LGLE in the macroscopic limit. We believe this result indicates that more work is necessary to determine for what observables and under what conditions, the phonon-confinement effects discussed in this section are significant or not.


\section{Conclusions}
    \label{sec:conclusions}
    In this paper we presented the lattice generalized Langevin equation, a model for simulating the effects of lattice phonons on surface atoms. The most important parameter in this model is the memory kernel. We parameterize the memory kernel using data from MD simulations, showing that it has a universal bimodal form due to coupling to both coherent acoustic oscillations as well as modes near the Debye frequency. This bimodal form is non-trivial, as it is not recovered in exactly solvable 1D systems with nearest neighbor interactions. Since the frequency of the acoustic oscillations depends on the size of the lattice, and nanoscale MD simulations impose unphysical phonon confinement effects, observables which depend surface phonons will also contain artifacts. We showed that that this was indeed the case for the surface trapping probability for a simple system of Argon on Pt(111). 

    The advantages of the LGLE model are, first, it's computational efficiency, as it reduces the the $N$ degrees of freedom of the lattice, to only a small handful on terms needed to describe the motion of a surface site. This dimensionality reduction makes the LGLE much faster than all-atom forcefields and particularly well-suited for the simulation of trajectory ensembles. Second, the insight that can be gained from studying the memory kernel, as we illustrated throughout this paper. Third, the transferability of the model. Once the LGLE is parameterized for a given type of lattice, any surface reaction with that lattice can use the same LGLE, given that the thermodynamic conditions (temperature/pressure/surface coverage/solvation) are roughly the same.

\section*{Supplementary Material}
See supplementary material for analysis of convergence of memory kernels using CF method, a comparison of memory kernels using CF and PO methods, memory kernels for other metal lattices other than Pt(111), memory kernels for 1D harmonic chains, and further details on scattering/desorption simulations.

\section*{Data Availability}
Data that support the findings of this study are available from the corresponding author upon reasonable request. The code used to process memory kernels and run GLE simulations is available on Github \url{https://github.com/afarahva/glepy}.  

\section*{Acknowledgements}
AF, MA, AAP, and APW were supported by the Office of Science of the U.S. Department of Energy under Contract No. DE-SC0019441.
This research used resources of the National Energy Research Scientific Computing Center, a DOE Office of Science User Facility supported by the Office of Science of the U.S. Department of Energy under Contract No.  DE-AC02-05CH11231. 
Ardavan Farahvash acknowledges support from the National Science Foundation Graduate Research Fellowship program. 

\bibliography{main.bib}

\providecommand{\latin}[1]{#1}
\makeatletter
\providecommand{\doi}
  {\begingroup\let\do\@makeother\dospecials
  \catcode`\{=1 \catcode`\}=2 \doi@aux}
\providecommand{\doi@aux}[1]{\endgroup\texttt{#1}}
\makeatother
\providecommand*\mcitethebibliography{\thebibliography}
\csname @ifundefined\endcsname{endmcitethebibliography}
  {\let\endmcitethebibliography\endthebibliography}{}
\begin{mcitethebibliography}{58}
\providecommand*\natexlab[1]{#1}
\providecommand*\mciteSetBstSublistMode[1]{}
\providecommand*\mciteSetBstMaxWidthForm[2]{}
\providecommand*\mciteBstWouldAddEndPuncttrue
  {\def\EndOfBibitem{\unskip.}}
\providecommand*\mciteBstWouldAddEndPunctfalse
  {\let\EndOfBibitem\relax}
\providecommand*\mciteSetBstMidEndSepPunct[3]{}
\providecommand*\mciteSetBstSublistLabelBeginEnd[3]{}
\providecommand*\EndOfBibitem{}
\mciteSetBstSublistMode{f}
\mciteSetBstMaxWidthForm{subitem}{(\alph{mcitesubitemcount})}
\mciteSetBstSublistLabelBeginEnd
  {\mcitemaxwidthsubitemform\space}
  {\relax}
  {\relax}

\bibitem[Tetenoire \latin{et~al.}(2022)Tetenoire, Ehlert, Juaristi, Saalfrank,
  and Alducin]{tetenoire_why_2022}
Tetenoire,~A.; Ehlert,~C.; Juaristi,~J.~I.; Saalfrank,~P.; Alducin,~M. Why
  {Ultrafast} {Photoinduced} {CO} {Desorption} {Dominates} over {Oxidation} on
  {Ru}(0001). \emph{The Journal of Physical Chemistry Letters} \textbf{2022},
  \emph{13}, 8516--8521\relax
\mciteBstWouldAddEndPuncttrue
\mciteSetBstMidEndSepPunct{\mcitedefaultmidpunct}
{\mcitedefaultendpunct}{\mcitedefaultseppunct}\relax
\EndOfBibitem
\bibitem[Alducin \latin{et~al.}(2019)Alducin, Camillone, Hong, and
  Juaristi]{alducin_electrons_2019}
Alducin,~M.; Camillone,~N.; Hong,~S.-Y.; Juaristi,~J.~I. Electrons and
  {Phonons} {Cooperate} in the {Laser}-{Induced} {Desorption} of {CO} from
  {Pd}(111). \emph{Physical Review Letters} \textbf{2019}, \emph{123},
  246802\relax
\mciteBstWouldAddEndPuncttrue
\mciteSetBstMidEndSepPunct{\mcitedefaultmidpunct}
{\mcitedefaultendpunct}{\mcitedefaultseppunct}\relax
\EndOfBibitem
\bibitem[Shakouri \latin{et~al.}(2018)Shakouri, Behler, Meyer, and
  Kroes]{shakouri_analysis_2018}
Shakouri,~K.; Behler,~J.; Meyer,~J.; Kroes,~G.-J. Analysis of {Energy}
  {Dissipation} {Channels} in a {Benchmark} {System} of {Activated}
  {Dissociation}: {N} $_{\textrm{2}}$ on {Ru}(0001). \emph{The Journal of
  Physical Chemistry C} \textbf{2018}, \emph{122}, 23470--23480\relax
\mciteBstWouldAddEndPuncttrue
\mciteSetBstMidEndSepPunct{\mcitedefaultmidpunct}
{\mcitedefaultendpunct}{\mcitedefaultseppunct}\relax
\EndOfBibitem
\bibitem[von Boehn \latin{et~al.}(2020)von Boehn, Foerster, von Boehn, Prat,
  Macià, Casals, Khaliq, Hernández-Mínguez, Aballe, and
  Imbihl]{von_boehn_promotion_2020}
von Boehn,~B.; Foerster,~M.; von Boehn,~M.; Prat,~J.; Macià,~F.; Casals,~B.;
  Khaliq,~M.~W.; Hernández-Mínguez,~A.; Aballe,~L.; Imbihl,~R. On the
  {Promotion} of {Catalytic} {Reactions} by {Surface} {Acoustic} {Waves}.
  \emph{Angewandte Chemie International Edition} \textbf{2020}, \emph{59},
  20224--20229\relax
\mciteBstWouldAddEndPuncttrue
\mciteSetBstMidEndSepPunct{\mcitedefaultmidpunct}
{\mcitedefaultendpunct}{\mcitedefaultseppunct}\relax
\EndOfBibitem
\bibitem[Stair and Weitz(1987)Stair, and
  Weitz]{stair_pulsed-laser-induced_1987}
Stair,~P.~C.; Weitz,~E. Pulsed-laser-induced desorption from metal surfaces.
  \emph{JOSA B} \textbf{1987}, \emph{4}, 255--260\relax
\mciteBstWouldAddEndPuncttrue
\mciteSetBstMidEndSepPunct{\mcitedefaultmidpunct}
{\mcitedefaultendpunct}{\mcitedefaultseppunct}\relax
\EndOfBibitem
\bibitem[Hall(1987)]{hall_pulsed-laser-induced_1987}
Hall,~R.~B. Pulsed-laser-induced desorption studies of the kinetics of surface
  reactions. \emph{The Journal of Physical Chemistry} \textbf{1987}, \emph{91},
  1007--1015\relax
\mciteBstWouldAddEndPuncttrue
\mciteSetBstMidEndSepPunct{\mcitedefaultmidpunct}
{\mcitedefaultendpunct}{\mcitedefaultseppunct}\relax
\EndOfBibitem
\bibitem[Bonn \latin{et~al.}(1999)Bonn, Funk, Hess, Denzler, Stampfl,
  Scheffler, Wolf, and Ertl]{bonn_phonon-_1999}
Bonn,~M.; Funk,~S.; Hess,~C.; Denzler,~D.~N.; Stampfl,~C.; Scheffler,~M.;
  Wolf,~M.; Ertl,~G. Phonon- {Versus} {Electron}-{Mediated} {Desorption} and
  {Oxidation} of {CO} on {Ru}(0001). \emph{Science} \textbf{1999}, \emph{285},
  1042--1045\relax
\mciteBstWouldAddEndPuncttrue
\mciteSetBstMidEndSepPunct{\mcitedefaultmidpunct}
{\mcitedefaultendpunct}{\mcitedefaultseppunct}\relax
\EndOfBibitem
\bibitem[Frischkorn and Wolf(2006)Frischkorn, and
  Wolf]{frischkorn_femtochemistry_2006}
Frischkorn,~C.; Wolf,~M. Femtochemistry at {Metal} {Surfaces}: {Nonadiabatic}
  {Reaction} {Dynamics}. \emph{Chemical Reviews} \textbf{2006}, \emph{106},
  4207--4233\relax
\mciteBstWouldAddEndPuncttrue
\mciteSetBstMidEndSepPunct{\mcitedefaultmidpunct}
{\mcitedefaultendpunct}{\mcitedefaultseppunct}\relax
\EndOfBibitem
\bibitem[Gladh \latin{et~al.}(2013)Gladh, Hansson, and
  Öström]{gladh_electron-_2013}
Gladh,~J.; Hansson,~T.; Öström,~H. Electron- and phonon-coupling in
  femtosecond laser-induced desorption of {CO} from {Ru}(0001). \emph{Surface
  Science} \textbf{2013}, \emph{615}, 65--71\relax
\mciteBstWouldAddEndPuncttrue
\mciteSetBstMidEndSepPunct{\mcitedefaultmidpunct}
{\mcitedefaultendpunct}{\mcitedefaultseppunct}\relax
\EndOfBibitem
\bibitem[Luntz and Bethune(1989)Luntz, and Bethune]{luntz_activation_1989}
Luntz,~A.~C.; Bethune,~D.~S. Activation of methane dissociation on a {Pt}(111)
  surface. \emph{The Journal of Chemical Physics} \textbf{1989}, \emph{90},
  1274--1280\relax
\mciteBstWouldAddEndPuncttrue
\mciteSetBstMidEndSepPunct{\mcitedefaultmidpunct}
{\mcitedefaultendpunct}{\mcitedefaultseppunct}\relax
\EndOfBibitem
\bibitem[Luntz and Harris(1991)Luntz, and Harris]{luntz_ch4_1991}
Luntz,~A.~C.; Harris,~J. {CH4} dissociation on metals: a quantum dynamics
  model. \emph{Surface Science} \textbf{1991}, \emph{258}, 397--426\relax
\mciteBstWouldAddEndPuncttrue
\mciteSetBstMidEndSepPunct{\mcitedefaultmidpunct}
{\mcitedefaultendpunct}{\mcitedefaultseppunct}\relax
\EndOfBibitem
\bibitem[Henkelman and Jónsson(2001)Henkelman, and
  Jónsson]{henkelman_theoretical_2001}
Henkelman,~G.; Jónsson,~H. Theoretical {Calculations} of {Dissociative}
  {Adsorption} of {CH} 4 on an {Ir}(111) {Surface}. \emph{Physical Review
  Letters} \textbf{2001}, \emph{86}, 664--667\relax
\mciteBstWouldAddEndPuncttrue
\mciteSetBstMidEndSepPunct{\mcitedefaultmidpunct}
{\mcitedefaultendpunct}{\mcitedefaultseppunct}\relax
\EndOfBibitem
\bibitem[Nave and Jackson(2007)Nave, and Jackson]{nave_methane_2007}
Nave,~S.; Jackson,~B. Methane {Dissociation} on {Ni}(111): {The} {Role} of
  {Lattice} {Reconstruction}. \emph{Physical Review Letters} \textbf{2007},
  \emph{98}, 173003, =\relax
\mciteBstWouldAddEndPuncttrue
\mciteSetBstMidEndSepPunct{\mcitedefaultmidpunct}
{\mcitedefaultendpunct}{\mcitedefaultseppunct}\relax
\EndOfBibitem
\bibitem[Nave and Jackson(2009)Nave, and Jackson]{nave_methane_2009}
Nave,~S.; Jackson,~B. Methane dissociation on {Ni}(111) and {Pt}(111):
  {Energetic} and dynamical studies. \emph{The Journal of Chemical Physics}
  \textbf{2009}, \emph{130}, 054701\relax
\mciteBstWouldAddEndPuncttrue
\mciteSetBstMidEndSepPunct{\mcitedefaultmidpunct}
{\mcitedefaultendpunct}{\mcitedefaultseppunct}\relax
\EndOfBibitem
\bibitem[Nave \latin{et~al.}(2010)Nave, Tiwari, and Jackson]{nave_methane_2010}
Nave,~S.; Tiwari,~A.~K.; Jackson,~B. Methane dissociation and adsorption on
  {Ni}(111), {Pt}(111), {Ni}(100), {Pt}(100), and {Pt}(110)-(1×2): {Energetic}
  study. \emph{The Journal of Chemical Physics} \textbf{2010}, \emph{132},
  054705, =\relax
\mciteBstWouldAddEndPuncttrue
\mciteSetBstMidEndSepPunct{\mcitedefaultmidpunct}
{\mcitedefaultendpunct}{\mcitedefaultseppunct}\relax
\EndOfBibitem
\bibitem[Tiwari \latin{et~al.}(2009)Tiwari, Nave, and
  Jackson]{tiwari_methane_2009}
Tiwari,~A.~K.; Nave,~S.; Jackson,~B. Methane {Dissociation} on {Ni}(111): {A}
  {New} {Understanding} of the {Lattice} {Effect}. \emph{Physical Review
  Letters} \textbf{2009}, \emph{103}, 253201, Publisher: American Physical
  Society\relax
\mciteBstWouldAddEndPuncttrue
\mciteSetBstMidEndSepPunct{\mcitedefaultmidpunct}
{\mcitedefaultendpunct}{\mcitedefaultseppunct}\relax
\EndOfBibitem
\bibitem[Tiwari \latin{et~al.}(2010)Tiwari, Nave, and
  Jackson]{tiwari_temperature_2010}
Tiwari,~A.~K.; Nave,~S.; Jackson,~B. The temperature dependence of methane
  dissociation on {Ni}(111) and {Pt}(111): {Mixed} quantum-classical studies of
  the lattice response. \emph{The Journal of Chemical Physics} \textbf{2010},
  \emph{132}, 134702\relax
\mciteBstWouldAddEndPuncttrue
\mciteSetBstMidEndSepPunct{\mcitedefaultmidpunct}
{\mcitedefaultendpunct}{\mcitedefaultseppunct}\relax
\EndOfBibitem
\bibitem[Egeberg \latin{et~al.}(2001)Egeberg, Larsen, and
  Chorkendorff]{egeberg_molecular_2001}
Egeberg,~R.~C.; Larsen,~J.~H.; Chorkendorff,~I. Molecular beam study of {N2}
  dissociation on {Ru}(0001). \emph{Physical Chemistry Chemical Physics}
  \textbf{2001}, \emph{3}, 2007--2011\relax
\mciteBstWouldAddEndPuncttrue
\mciteSetBstMidEndSepPunct{\mcitedefaultmidpunct}
{\mcitedefaultendpunct}{\mcitedefaultseppunct}\relax
\EndOfBibitem
\bibitem[Diekhöner \latin{et~al.}(2001)Diekhöner, Mortensen, Baurichter,
  Jensen, Petrunin, and Luntz]{diekhoner_n2_2001}
Diekhöner,~L.; Mortensen,~H.; Baurichter,~A.; Jensen,~E.; Petrunin,~V.~V.;
  Luntz,~A.~C. N2 dissociative adsorption on {Ru}(0001): {The} role of energy
  loss. \emph{The Journal of Chemical Physics} \textbf{2001}, \emph{115},
  9028--9035\relax
\mciteBstWouldAddEndPuncttrue
\mciteSetBstMidEndSepPunct{\mcitedefaultmidpunct}
{\mcitedefaultendpunct}{\mcitedefaultseppunct}\relax
\EndOfBibitem
\bibitem[Nattino \latin{et~al.}(2015)Nattino, Costanzo, and
  Kroes]{nattino_n2_2015}
Nattino,~F.; Costanzo,~F.; Kroes,~G.-J. N2 dissociation on {W}(110): {An} ab
  initio molecular dynamics study on the effect of phonons. \emph{The Journal
  of Chemical Physics} \textbf{2015}, \emph{142}, 104702\relax
\mciteBstWouldAddEndPuncttrue
\mciteSetBstMidEndSepPunct{\mcitedefaultmidpunct}
{\mcitedefaultendpunct}{\mcitedefaultseppunct}\relax
\EndOfBibitem
\bibitem[Shakouri \latin{et~al.}(2017)Shakouri, Behler, Meyer, and
  Kroes]{shakouri_accurate_2017}
Shakouri,~K.; Behler,~J.; Meyer,~J.; Kroes,~G.-J. Accurate {Neural} {Network}
  {Description} of {Surface} {Phonons} in {Reactive} {Gas}–{Surface}
  {Dynamics}: {N2} + {Ru}(0001). \emph{The Journal of Physical Chemistry
  Letters} \textbf{2017}, \emph{8}, 2131--2136\relax
\mciteBstWouldAddEndPuncttrue
\mciteSetBstMidEndSepPunct{\mcitedefaultmidpunct}
{\mcitedefaultendpunct}{\mcitedefaultseppunct}\relax
\EndOfBibitem
\bibitem[Inoue(2007)]{inoue_effects_2007}
Inoue,~Y. Effects of acoustic waves-induced dynamic lattice distortion on
  catalytic and adsorptive properties of metal, alloy and metal oxide surfaces.
  \emph{Surface Science Reports} \textbf{2007}, \emph{62}, 305--336\relax
\mciteBstWouldAddEndPuncttrue
\mciteSetBstMidEndSepPunct{\mcitedefaultmidpunct}
{\mcitedefaultendpunct}{\mcitedefaultseppunct}\relax
\EndOfBibitem
\bibitem[Adelman and Doll(1976)Adelman, and Doll]{adelman_generalized_1976}
Adelman,~S.~A.; Doll,~J.~D. Generalized {Langevin} equation approach for
  atom/solid‐surface scattering: {General} formulation for classical
  scattering off harmonic solids. \emph{The Journal of Chemical Physics}
  \textbf{1976}, \emph{64}, 2375--2388\relax
\mciteBstWouldAddEndPuncttrue
\mciteSetBstMidEndSepPunct{\mcitedefaultmidpunct}
{\mcitedefaultendpunct}{\mcitedefaultseppunct}\relax
\EndOfBibitem
\bibitem[Tully(1980)]{tully_dynamics_1980}
Tully,~J.~C. Dynamics of gas–surface interactions: {3D} generalized
  {Langevin} model applied to fcc and bcc surfaces. \emph{The Journal of
  Chemical Physics} \textbf{1980}, \emph{73}, 1975--1985\relax
\mciteBstWouldAddEndPuncttrue
\mciteSetBstMidEndSepPunct{\mcitedefaultmidpunct}
{\mcitedefaultendpunct}{\mcitedefaultseppunct}\relax
\EndOfBibitem
\bibitem[Busnengo \latin{et~al.}(2004)Busnengo, Dong, and
  Salin]{busnengo_trapping_2004}
Busnengo,~H.~F.; Dong,~W.; Salin,~A. Trapping, {Molecular} {Adsorption}, and
  {Precursors} for {Nonactivated} {Chemisorption}. \emph{Physical Review
  Letters} \textbf{2004}, \emph{93}, 236103, =\relax
\mciteBstWouldAddEndPuncttrue
\mciteSetBstMidEndSepPunct{\mcitedefaultmidpunct}
{\mcitedefaultendpunct}{\mcitedefaultseppunct}\relax
\EndOfBibitem
\bibitem[Bukas \latin{et~al.}(2015)Bukas, Mitra, Meyer, and
  Reuter]{bukas_fingerprints_2015}
Bukas,~V.~J.; Mitra,~S.; Meyer,~J.; Reuter,~K. Fingerprints of energy
  dissipation for exothermic surface chemical reactions: {O2} on {Pd}(100).
  \emph{The Journal of Chemical Physics} \textbf{2015}, \emph{143},
  034705\relax
\mciteBstWouldAddEndPuncttrue
\mciteSetBstMidEndSepPunct{\mcitedefaultmidpunct}
{\mcitedefaultendpunct}{\mcitedefaultseppunct}\relax
\EndOfBibitem
\bibitem[Nattino \latin{et~al.}(2016)Nattino, Galparsoro, Costanzo,
  Díez~Muiño, Alducin, and Kroes]{nattino_modeling_2016}
Nattino,~F.; Galparsoro,~O.; Costanzo,~F.; Díez~Muiño,~R.; Alducin,~M.;
  Kroes,~G.-J. Modeling surface motion effects in {N2} dissociation on
  {W}(110): {Ab} initio molecular dynamics calculations and generalized
  {Langevin} oscillator model. \emph{The Journal of Chemical Physics}
  \textbf{2016}, \emph{144}, 244708\relax
\mciteBstWouldAddEndPuncttrue
\mciteSetBstMidEndSepPunct{\mcitedefaultmidpunct}
{\mcitedefaultendpunct}{\mcitedefaultseppunct}\relax
\EndOfBibitem
\bibitem[Zhou \latin{et~al.}(2020)Zhou, Zhou, Hu, and Xie]{zhou_dynamics_2020}
Zhou,~Y.; Zhou,~L.; Hu,~X.; Xie,~D. Dynamics {Studies} of {O2} {Collision} on
  {Pt}(111) {Using} a {Global} {Potential} {Energy} {Surface}. \emph{The
  Journal of Physical Chemistry C} \textbf{2020}, \emph{124},
  10573--10583\relax
\mciteBstWouldAddEndPuncttrue
\mciteSetBstMidEndSepPunct{\mcitedefaultmidpunct}
{\mcitedefaultendpunct}{\mcitedefaultseppunct}\relax
\EndOfBibitem
\bibitem[Zhou and Jiang(2019)Zhou, and Jiang]{zhou_modified_2019}
Zhou,~X.; Jiang,~B. A modified generalized {Langevin} oscillator model for
  activated gas-surface reactions. \emph{The Journal of Chemical Physics}
  \textbf{2019}, \emph{150}, 024704\relax
\mciteBstWouldAddEndPuncttrue
\mciteSetBstMidEndSepPunct{\mcitedefaultmidpunct}
{\mcitedefaultendpunct}{\mcitedefaultseppunct}\relax
\EndOfBibitem
\bibitem[Canales and Sesé(1998)Canales, and Sesé]{canales_generalized_1998}
Canales,~M.; Sesé,~G. Generalized {Langevin} dynamics simulations of {NaCl}
  electrolyte solutions. \emph{The Journal of Chemical Physics} \textbf{1998},
  \emph{109}, 6004--6011\relax
\mciteBstWouldAddEndPuncttrue
\mciteSetBstMidEndSepPunct{\mcitedefaultmidpunct}
{\mcitedefaultendpunct}{\mcitedefaultseppunct}\relax
\EndOfBibitem
\bibitem[Gordon \latin{et~al.}(2009)Gordon, Krishnamurthy, and
  Chung]{gordon_generalized_2009}
Gordon,~D.; Krishnamurthy,~V.; Chung,~S.-H. Generalized {Langevin} models of
  molecular dynamics simulations with applications to ion channels. \emph{The
  Journal of Chemical Physics} \textbf{2009}, \emph{131}, 134102\relax
\mciteBstWouldAddEndPuncttrue
\mciteSetBstMidEndSepPunct{\mcitedefaultmidpunct}
{\mcitedefaultendpunct}{\mcitedefaultseppunct}\relax
\EndOfBibitem
\bibitem[Jung \latin{et~al.}(2017)Jung, Hanke, and Schmid]{jung_iterative_2017}
Jung,~G.; Hanke,~M.; Schmid,~F. Iterative {Reconstruction} of {Memory}
  {Kernels}. \emph{Journal of Chemical Theory and Computation} \textbf{2017},
  \emph{13}, 2481--2488\relax
\mciteBstWouldAddEndPuncttrue
\mciteSetBstMidEndSepPunct{\mcitedefaultmidpunct}
{\mcitedefaultendpunct}{\mcitedefaultseppunct}\relax
\EndOfBibitem
\bibitem[Li \latin{et~al.}(2017)Li, Lee, Darve, and
  Karniadakis]{li_computing_2017}
Li,~Z.; Lee,~H.~S.; Darve,~E.; Karniadakis,~G.~E. Computing the non-{Markovian}
  coarse-grained interactions derived from the {Mori}–{Zwanzig} formalism in
  molecular systems: {Application} to polymer melts. \emph{The Journal of
  Chemical Physics} \textbf{2017}, \emph{146}, 014104\relax
\mciteBstWouldAddEndPuncttrue
\mciteSetBstMidEndSepPunct{\mcitedefaultmidpunct}
{\mcitedefaultendpunct}{\mcitedefaultseppunct}\relax
\EndOfBibitem
\bibitem[Wang \latin{et~al.}(2019)Wang, Li, and
  Pan]{wang_implicit-solvent_2019}
Wang,~S.; Li,~Z.; Pan,~W. Implicit-solvent coarse-grained modeling for polymer
  solutions via {Mori}-{Zwanzig} formalism. \emph{Soft Matter} \textbf{2019},
  \emph{15}, 7567--7582\relax
\mciteBstWouldAddEndPuncttrue
\mciteSetBstMidEndSepPunct{\mcitedefaultmidpunct}
{\mcitedefaultendpunct}{\mcitedefaultseppunct}\relax
\EndOfBibitem
\bibitem[Hänggi(1997)]{hanggi_generalized_1997}
Hänggi,~P. Generalized langevin equations: {A} useful tool for the perplexed
  modeller of nonequilibrium fluctuations? Stochastic {Dynamics}. Berlin,
  Heidelberg, 1997; pp 15--22\relax
\mciteBstWouldAddEndPuncttrue
\mciteSetBstMidEndSepPunct{\mcitedefaultmidpunct}
{\mcitedefaultendpunct}{\mcitedefaultseppunct}\relax
\EndOfBibitem
\bibitem[Berne and Harp(1970)Berne, and Harp]{berne_calculation_1970}
Berne,~B.~J.; Harp,~G.~D. \emph{Advances in {Chemical} {Physics}}; John Wiley
  \& Sons, Ltd, 1970; pp 63--227\relax
\mciteBstWouldAddEndPuncttrue
\mciteSetBstMidEndSepPunct{\mcitedefaultmidpunct}
{\mcitedefaultendpunct}{\mcitedefaultseppunct}\relax
\EndOfBibitem
\bibitem[Ceriotti \latin{et~al.}(2010)Ceriotti, Bussi, and
  Parrinello]{ceriotti_colored-noise_2010}
Ceriotti,~M.; Bussi,~G.; Parrinello,~M. Colored-{Noise} {Thermostats} à la
  {Carte}. \emph{Journal of Chemical Theory and Computation} \textbf{2010},
  \emph{6}, 1170--1180\relax
\mciteBstWouldAddEndPuncttrue
\mciteSetBstMidEndSepPunct{\mcitedefaultmidpunct}
{\mcitedefaultendpunct}{\mcitedefaultseppunct}\relax
\EndOfBibitem
\bibitem[Baczewski and Bond(2013)Baczewski, and Bond]{baczewski_numerical_2013}
Baczewski,~A.~D.; Bond,~S.~D. Numerical {Integration} of the {Extended}
  {Variable} {Generalized} {Langevin} {Equation} with a {Positive} {Prony}
  {Representable} {Memory} {Kernel}. \emph{The Journal of Chemical Physics}
  \textbf{2013}, \emph{139}, 044107\relax
\mciteBstWouldAddEndPuncttrue
\mciteSetBstMidEndSepPunct{\mcitedefaultmidpunct}
{\mcitedefaultendpunct}{\mcitedefaultseppunct}\relax
\EndOfBibitem
\bibitem[N{\o}rskov and Lang(1980)N{\o}rskov, and Lang]{Norskov1980}
N{\o}rskov,~J.~K.; Lang,~N.~D. Effective-medium theory of chemical binding:
  Application to chemisorption. \emph{Physical Review B} \textbf{1980},
  \emph{21}, 2131--2136\relax
\mciteBstWouldAddEndPuncttrue
\mciteSetBstMidEndSepPunct{\mcitedefaultmidpunct}
{\mcitedefaultendpunct}{\mcitedefaultseppunct}\relax
\EndOfBibitem
\bibitem[Daw and Baskes(1983)Daw, and Baskes]{Daw1983}
Daw,~M.~S.; Baskes,~M.~I. Semiempirical, Quantum Mechanical Calculation of
  Hydrogen Embrittlement in Metals. \emph{Physical Review Letters}
  \textbf{1983}, \emph{50}, 1285--1288\relax
\mciteBstWouldAddEndPuncttrue
\mciteSetBstMidEndSepPunct{\mcitedefaultmidpunct}
{\mcitedefaultendpunct}{\mcitedefaultseppunct}\relax
\EndOfBibitem
\bibitem[Larsen \latin{et~al.}(2017)Larsen, Mortensen, Blomqvist, Castelli,
  Christensen, Dułak, Friis, Groves, Hammer, Hargus, Hermes, Jennings, Jensen,
  Kermode, Kitchin, Kolsbjerg, Kubal, Kaasbjerg, Lysgaard, Maronsson, Maxson,
  Olsen, Pastewka, Peterson, Rostgaard, Schiøtz, Schütt, Strange, Thygesen,
  Vegge, Vilhelmsen, Walter, Zeng, and Jacobsen]{larsen_atomic_2017}
Larsen,~A.~H. \latin{et~al.}  The atomic simulation environment—a {Python}
  library for working with atoms. \emph{Journal of Physics: Condensed Matter}
  \textbf{2017}, \emph{29}, 273002\relax
\mciteBstWouldAddEndPuncttrue
\mciteSetBstMidEndSepPunct{\mcitedefaultmidpunct}
{\mcitedefaultendpunct}{\mcitedefaultseppunct}\relax
\EndOfBibitem
\bibitem[Tadmor \latin{et~al.}(2011)Tadmor, Elliott, Sethna, Miller, and
  Becker]{tadmor:elliott:2011}
Tadmor,~E.~B.; Elliott,~R.~S.; Sethna,~J.~P.; Miller,~R.~E.; Becker,~C.~A. The
  potential of atomistic simulations and the {K}nowledgebase of {I}nteratomic
  {M}odels. \emph{{JOM}} \textbf{2011}, \emph{63}, 17\relax
\mciteBstWouldAddEndPuncttrue
\mciteSetBstMidEndSepPunct{\mcitedefaultmidpunct}
{\mcitedefaultendpunct}{\mcitedefaultseppunct}\relax
\EndOfBibitem
\bibitem[Schiøtz(2019)]{OpenKIM-MO:108408461881:001}
Schiøtz,~J. {EMT} potential for {N}i developed by {J}acobsen, {S}tolze, and
  {N}orskov (1996) v001. OpenKIM, \url{https://doi.org/10.25950/ef22cbc3},
  2019\relax
\mciteBstWouldAddEndPuncttrue
\mciteSetBstMidEndSepPunct{\mcitedefaultmidpunct}
{\mcitedefaultendpunct}{\mcitedefaultseppunct}\relax
\EndOfBibitem
\bibitem[Schiøtz(2019)]{OpenKIM-MD:128315414717:004}
Schiøtz,~J. {E}ffective {M}edium {T}heory ({EMT}) potential driver v004.
  OpenKIM, \url{https://doi.org/10.25950/7e5b8be7}, 2019\relax
\mciteBstWouldAddEndPuncttrue
\mciteSetBstMidEndSepPunct{\mcitedefaultmidpunct}
{\mcitedefaultendpunct}{\mcitedefaultseppunct}\relax
\EndOfBibitem
\bibitem[Tadmor(2018)]{OpenKIM-MO:757342646688:000}
Tadmor,~E. {EAM} potential ({LAMMPS} cubic hermite tabulation) for {P}t
  ({U}niversal3) developed by {F}oiles, {B}askes, and {D}aw (1986) v000.
  OpenKIM, \url{https://doi.org/10.25950/24de6537}, 2018\relax
\mciteBstWouldAddEndPuncttrue
\mciteSetBstMidEndSepPunct{\mcitedefaultmidpunct}
{\mcitedefaultendpunct}{\mcitedefaultseppunct}\relax
\EndOfBibitem
\bibitem[Elliott(2018)]{OpenKIM-MD:120291908751:005}
Elliott,~R.~S. {EAM} {M}odel {D}river for tabulated potentials with cubic
  {H}ermite spline interpolation as used in {LAMMPS} v005. OpenKIM,
  \url{https://doi.org/10.25950/68defa36}, 2018\relax
\mciteBstWouldAddEndPuncttrue
\mciteSetBstMidEndSepPunct{\mcitedefaultmidpunct}
{\mcitedefaultendpunct}{\mcitedefaultseppunct}\relax
\EndOfBibitem
\bibitem[Jacobsen \latin{et~al.}(1996)Jacobsen, Stoltze, and
  Nørskov]{jacobsen_semi-empirical_1996}
Jacobsen,~K.~W.; Stoltze,~P.; Nørskov,~J.~K. A semi-empirical effective medium
  theory for metals and alloys. \emph{Surface Science} \textbf{1996},
  \emph{366}, 394--402\relax
\mciteBstWouldAddEndPuncttrue
\mciteSetBstMidEndSepPunct{\mcitedefaultmidpunct}
{\mcitedefaultendpunct}{\mcitedefaultseppunct}\relax
\EndOfBibitem
\bibitem[Zhou \latin{et~al.}(2004)Zhou, Johnson, and
  Wadley]{zhou_misfit-energy-increasing_2004}
Zhou,~X.~W.; Johnson,~R.~A.; Wadley,~H. N.~G. Misfit-energy-increasing
  dislocations in vapor-deposited {CoFe}/{NiFe} multilayers. \emph{Physical
  Review B} \textbf{2004}, \emph{69}, 144113\relax
\mciteBstWouldAddEndPuncttrue
\mciteSetBstMidEndSepPunct{\mcitedefaultmidpunct}
{\mcitedefaultendpunct}{\mcitedefaultseppunct}\relax
\EndOfBibitem
\bibitem[Heinz \latin{et~al.}(2008)Heinz, Vaia, Farmer, and
  Naik]{heinz_accurate_2008}
Heinz,~H.; Vaia,~R.~A.; Farmer,~B.~L.; Naik,~R.~R. Accurate {Simulation} of
  {Surfaces} and {Interfaces} of {Face}-{Centered} {Cubic} {Metals} {Using}
  12-6 and 9-6 {Lennard}-{Jones} {Potentials}. \emph{The Journal of Physical
  Chemistry C} \textbf{2008}, \emph{112}, 17281--17290\relax
\mciteBstWouldAddEndPuncttrue
\mciteSetBstMidEndSepPunct{\mcitedefaultmidpunct}
{\mcitedefaultendpunct}{\mcitedefaultseppunct}\relax
\EndOfBibitem
\bibitem[Berendsen \latin{et~al.}(1987)Berendsen, Grigera, and
  Straatsma]{berendsen_missing_1987}
Berendsen,~H. J.~C.; Grigera,~J.~R.; Straatsma,~T.~P. The missing term in
  effective pair potentials. \emph{The Journal of Physical Chemistry}
  \textbf{1987}, \emph{91}, 6269--6271\relax
\mciteBstWouldAddEndPuncttrue
\mciteSetBstMidEndSepPunct{\mcitedefaultmidpunct}
{\mcitedefaultendpunct}{\mcitedefaultseppunct}\relax
\EndOfBibitem
\bibitem[Florencio and Lee(1985)Florencio, and Lee]{florencio_exact_1985}
Florencio,~J.; Lee,~M.~H. Exact time evolution of a classical
  harmonic-oscillator chain. \emph{Physical Review A} \textbf{1985}, \emph{31},
  3231--3236\relax
\mciteBstWouldAddEndPuncttrue
\mciteSetBstMidEndSepPunct{\mcitedefaultmidpunct}
{\mcitedefaultendpunct}{\mcitedefaultseppunct}\relax
\EndOfBibitem
\bibitem[Lee(2016)]{lee_local_2016}
Lee,~M.~H. Local {Dynamics} in an {Infinite} {Harmonic} {Chain}.
  \emph{Symmetry} \textbf{2016}, \emph{8}, 22\relax
\mciteBstWouldAddEndPuncttrue
\mciteSetBstMidEndSepPunct{\mcitedefaultmidpunct}
{\mcitedefaultendpunct}{\mcitedefaultseppunct}\relax
\EndOfBibitem
\bibitem[Arbouet \latin{et~al.}(2003)Arbouet, Voisin, Christofilos, Langot,
  Fatti, Vallée, Lermé, Celep, Cottancin, Gaudry, Pellarin, Broyer, Maillard,
  Pileni, and Treguer]{arbouet_electron-phonon_2003}
Arbouet,~A.; Voisin,~C.; Christofilos,~D.; Langot,~P.; Fatti,~N.~D.;
  Vallée,~F.; Lermé,~J.; Celep,~G.; Cottancin,~E.; Gaudry,~M.; Pellarin,~M.;
  Broyer,~M.; Maillard,~M.; Pileni,~M.~P.; Treguer,~M. Electron-{Phonon}
  {Scattering} in {Metal} {Clusters}. \emph{Physical Review Letters}
  \textbf{2003}, \emph{90}, 177401\relax
\mciteBstWouldAddEndPuncttrue
\mciteSetBstMidEndSepPunct{\mcitedefaultmidpunct}
{\mcitedefaultendpunct}{\mcitedefaultseppunct}\relax
\EndOfBibitem
\bibitem[Sitz(2002)]{sitz_gas_2002}
Sitz,~G.~O. Gas surface interactions studied with state-prepared molecules.
  \emph{Reports on Progress in Physics} \textbf{2002}, \emph{65}, 1165\relax
\mciteBstWouldAddEndPuncttrue
\mciteSetBstMidEndSepPunct{\mcitedefaultmidpunct}
{\mcitedefaultendpunct}{\mcitedefaultseppunct}\relax
\EndOfBibitem
\bibitem[Juurlink \latin{et~al.}(2009)Juurlink, Killelea, and
  Utz]{juurlink_state-resolved_2009}
Juurlink,~L. B.~F.; Killelea,~D.~R.; Utz,~A.~L. State-resolved probes of
  methane dissociation dynamics. \emph{Progress in Surface Science}
  \textbf{2009}, \emph{84}, 69--134\relax
\mciteBstWouldAddEndPuncttrue
\mciteSetBstMidEndSepPunct{\mcitedefaultmidpunct}
{\mcitedefaultendpunct}{\mcitedefaultseppunct}\relax
\EndOfBibitem
\bibitem[Chadwick and Beck(2017)Chadwick, and Beck]{chadwick_quantum_2017}
Chadwick,~H.; Beck,~R.~D. Quantum {State}-{Resolved} {Studies} of
  {Chemisorption} {Reactions}. \emph{Annual Review of Physical Chemistry}
  \textbf{2017}, \emph{68}, 39--61\relax
\mciteBstWouldAddEndPuncttrue
\mciteSetBstMidEndSepPunct{\mcitedefaultmidpunct}
{\mcitedefaultendpunct}{\mcitedefaultseppunct}\relax
\EndOfBibitem
\bibitem[Chen \latin{et~al.}(2012)Chen, Al-Saidi, and
  Karl~Johnson]{chen_role_2012}
Chen,~D.-L.; Al-Saidi,~W.~A.; Karl~Johnson,~J. The role of van der {Waals}
  interactions in the adsorption of noble gases on metal surfaces.
  \emph{Journal of Physics: Condensed Matter} \textbf{2012}, \emph{24},
  424211\relax
\mciteBstWouldAddEndPuncttrue
\mciteSetBstMidEndSepPunct{\mcitedefaultmidpunct}
{\mcitedefaultendpunct}{\mcitedefaultseppunct}\relax
\EndOfBibitem
\end{mcitethebibliography}

\end{document}


\title[\\]{Supplementary information for "On using the generalized Langevin equation to model substrate phonons and their role in surface adsorption and desorption"}

\author{Ardavan Farahvash}
\affiliation{Department of Chemistry, Massachusetts Institute of Technology, Cambridge, Massachusetts 02139, USA}

\author{Mayank Agrawal}
\affiliation{Department of Chemical Engineering, Brown University, Providence, Rhode Island 02912, USA}

\author{Andrew Peterson}
\affiliation{Department of Chemical Engineering, Brown University, Providence, Rhode Island 02912, USA}

\author{Adam P. Willard}
\email{apwillard@mit.edu}
\affiliation{Department of Chemistry, Massachusetts Institute of Technology, Cambridge, Massachusetts 02139, USA}

\maketitle 

\section{Convergence of CF memory kernel}
    \label{sec:convergence}
    
\begin{figure}[h]  
       \centering
       \includegraphics*[width=6.25in]{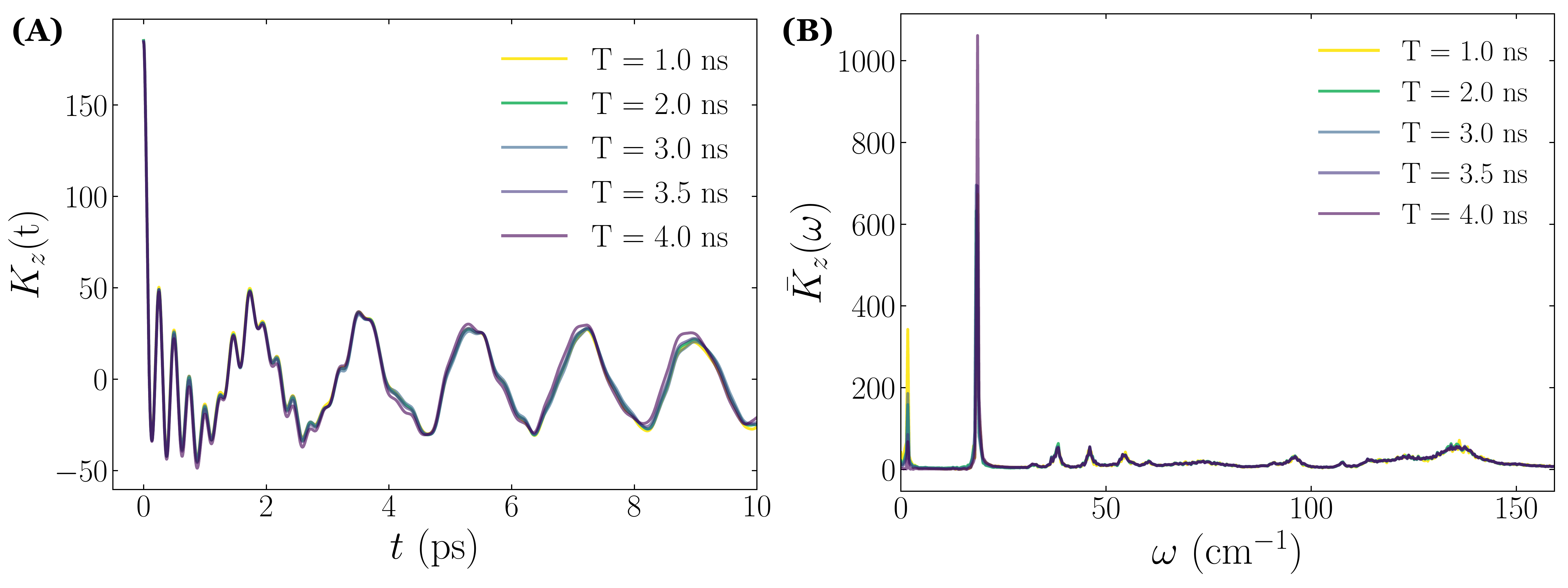}
       \caption{Convergence of memory kernel (A) and power spectra (B) for 4x4x4 Pt(111) lattice taken as a function of the simulation length $T$}
       \label{figure_s1}
    \end{figure}

\begin{figure}[h]  
       \centering
       \includegraphics*[width=6.25in]{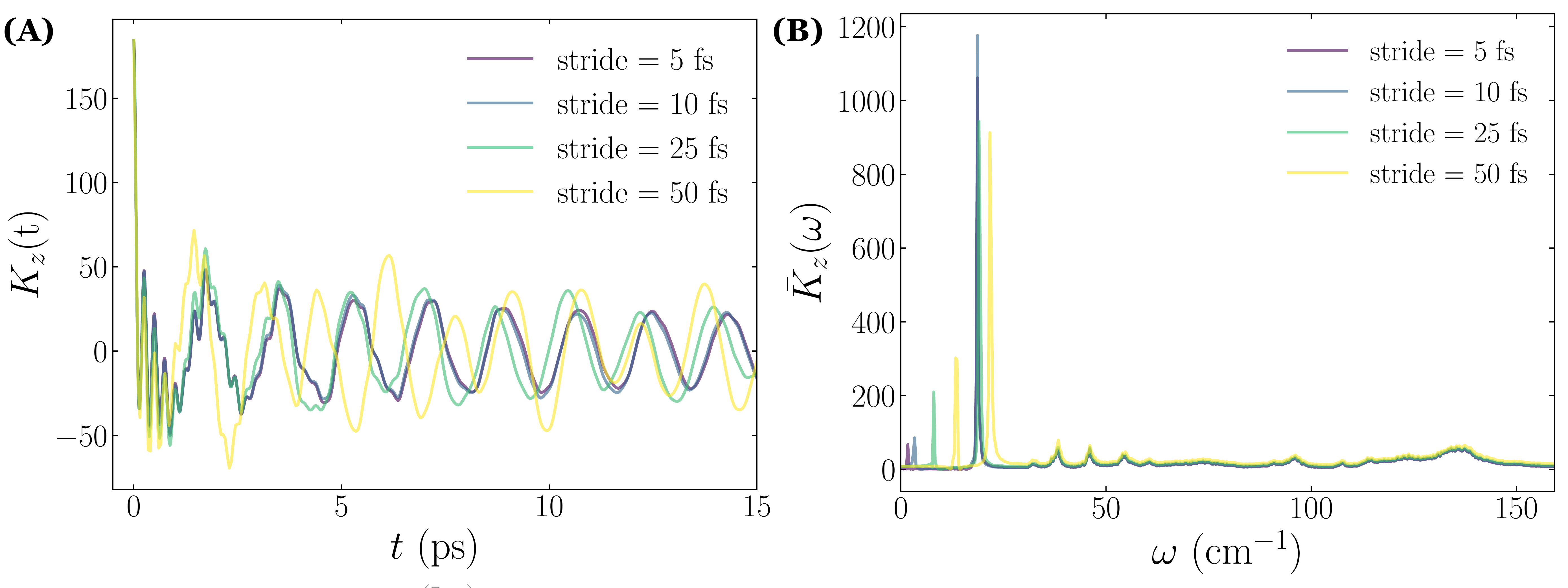}
       \caption{Convergence of memory kernel (A) and power spectra (B) for 4x4x4 Pt(111) lattice taken as a function of the step length.}
       \label{figure_s2}
    \end{figure}

    We have measured the convergence of the memory kernel and force power spectra both in terms of the step size (stride) between simulation snapshots used to compute the time correlation functions in Eq.9 and in terms of the total length the simulation used to compute the time correlation functions $T$. In Figure \ref{figure_s1} we illustrate the simulation length convergence. We see that the short time ($ < 10 \text{ps}$ or $> 30 cm^{-1}$) statistics converge very quickly with respect to simulation length. The long time, low frequency statistics however are slower to converge, especially in terms of the heights of the associated peaks. In particular, using simulation lengths less than 2ns seems to produce artificial oscillations at a very low frequency ($\sim 3 \text{cm}^{\text{-1}}$) in the memory kernel.
    In terms of stride, Figure~\ref{figure_s2} shows that the memory kernel is very well converged using step size of 10fs between snapshots. 

    \FloatBarrier

\section{Memory Kernel Fitting}
    \label{sec:Fitting}
\begin{figure}[h]  
       \centering
       \includegraphics*[width=6.25in]{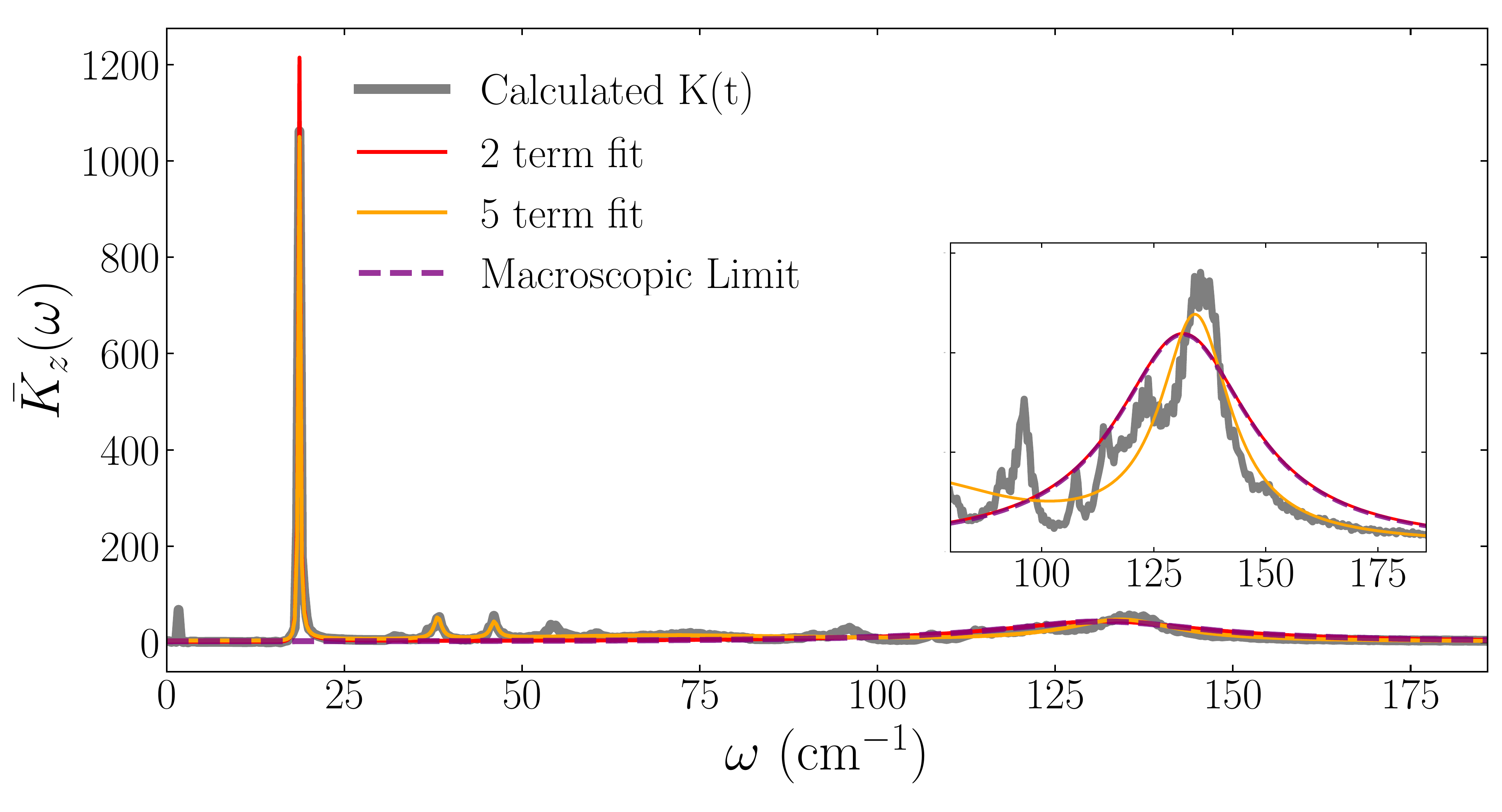}
       \caption{Power spectra for a 4x4x4 Pt(111) lattice calculated using the CF method overlayed with a 2 term Lorentzian fits, 5 term Lorentzian fit, and the macrscopic limit corresponding to only fitting the region of the power spectra near the Debye frequency. }
       \label{figure_s3}
    \end{figure}

    Parameters for memory kernels fit and used in simulations in the main text are provided in Table \ref{tab:gle} and depicted in Figure~\ref{figure_s3}.
    
    \begin{table}[htb]
        \caption{Parameters for memory kernel fit to a sum of exponentially damped cosines.}
        \label{tab:gle}
        \begin{tabularx}{\textwidth}{XXXX}
            \toprule
            & $C$ ($\mathrm{ps^{-1}}$) & $\gamma$ ($\mathrm{ps^{-1}}$) & $\omega$ ($\mathrm{cm^{-1}}$)  \\ 
            \midrule
            \textbf{2 term} & 146.3 & 3.42 & 131.5 \\
                            & 38.0  & 0.03 & 18.6 \\
            \midrule
            \textbf{5 term} & 79.4 & 1.78 & 134.2 \\
                            & 65.3 & 5.24 & 69.6 \\
                            & 32.5 & 0.03 & 18.6 \\
                            & 5.2  & 0.12 & 38.1 \\
                            & 4.0  & 0.12 & 46.1 \\
            \midrule
            \textbf{Macroscopic Limit} & 146.3 & 3.42 & 131.5\\
            \bottomrule
        \end{tabularx}
    \end{table} 

    \FloatBarrier
    \clearpage

\section{Comparison of memory kernels using PO and CF techniques}
    \label{sec:PO_comparison}
    In Figure \ref{figure_s3} we compare the memory kernel and power spectra computed using correlation function method and projection operator method detailed in Section 2 of the main text. Two differences of note are the small frequency shift between the two methods, and that the CF memory kernel is much smoother.  As we verified in Figures~\ref{figure_s1} and \ref{figure_s2} that the memory kernel is well converged, the difference between the PO curves and CF curves in Figure~\ref{figure_s3} can only be attributed to small anharmonicities in the EMT forcefield. Despite some differences between the two methods, we see that the PO method gives the same bimodal behavior as the CF approach. 
    \begin{figure}[htb]  
       \centering
       \includegraphics*[width=6.25in]{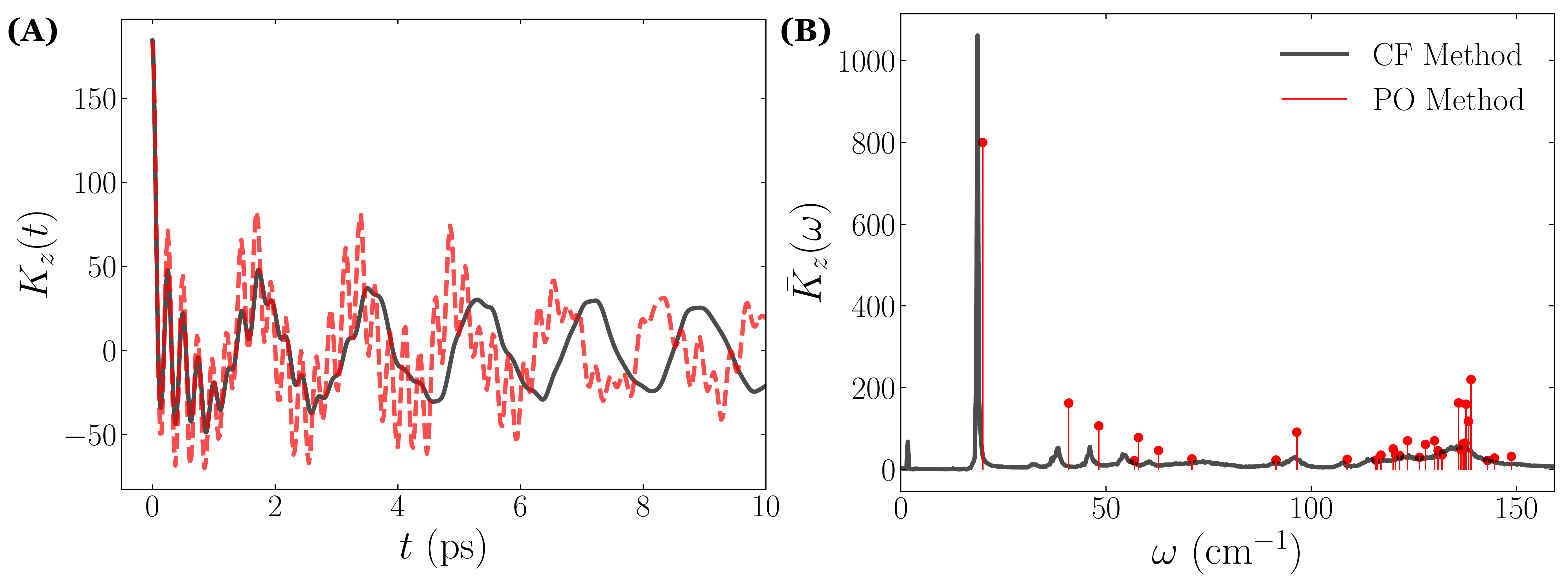}
       \caption{Comparison of memory kernel (A) and power spectra (B) using CF and PO methods for 4x4x4 Pt(111) lattice. The $\delta$ function form of the PO spectral density is represented with vertical lines.}
       \label{figure_s4}
    \end{figure}
    \clearpage

\section{Macroscopic Limit of 1D Harmonic Chain}
    \label{sec:1D_chain}
    Consider a 1D chain of harmonic oscillators with spring constant $k=m\omega^2$ and periodic boundary conditions. Every site in the chain is identical, and the dynamical matrix is given by,
    \begin{equation}
        \mathbf{D}^2 = 
        \omega^2
        \begin{pmatrix}
        -2 & 1 & 0 & 0 & \ldots & -1 \\
        -1 & 2 & -1 & 0 & \ldots & 0 \\
        \vdots &  & \ddots & \ddots & & \vdots \\
        0 & \ldots & 0 & -1 & 2 & -1 \\
        -1 & \ldots & 0 & 0 & -1 & 2 \\
        \end{pmatrix}.
    \end{equation}
    Taking our system to be a single site in the lattice, the resulting bath projected matrix is given by
    \begin{equation}
    \mathbf{D}^2_{QQ} = 
    \omega^2
    \begin{pmatrix}
         2 & -1 & 0 & \ldots & 0 \\
        -1 & 2 & -1 & \ldots & 0 \\
        \vdots &   & \ddots & & \vdots \\
        0 & \ldots & -1 & 2 & -1 \\
        0 & \ldots & 0 & -1 & 2 \\
    \end{pmatrix}.
    \end{equation}
    This matrix may be diagonalized analytically allowing one to find a solution to the memory kernel via Eq.5,
    \begin{equation}
        K(t) = \frac{4 k}{N} \sum_{n=1}^N \cos^2(\theta_n) \cos \left(2 \omega t \sin(\theta_n) \right),
    \end{equation}
    where $N$ is the total length of the chain and $\theta_n = \frac{n \pi}{2(N + 1)}$. If we take the limit as $N \rightarrow \infty$, we see that this sum converges to an integral, 
    \begin{equation}
        K(t) = 8 \pi k \int_0^{\pi/2} d\theta \cos^2(\theta) \cos \left(2 \omega t \sin(\theta) \right).
    \end{equation}
    This integral has no closed form solution. However it can be expressed in terms of Bessel functions,
    \begin{equation}
        K(t) =  \frac{4 \omega^2}{\pi} \frac{J_1(2 \omega t)}{t},
    \end{equation}
    where $J$ is a Bessel function of the first kind.
    
    \begin{figure}[htb]  
       \centering
       \includegraphics*[width=6.25in]{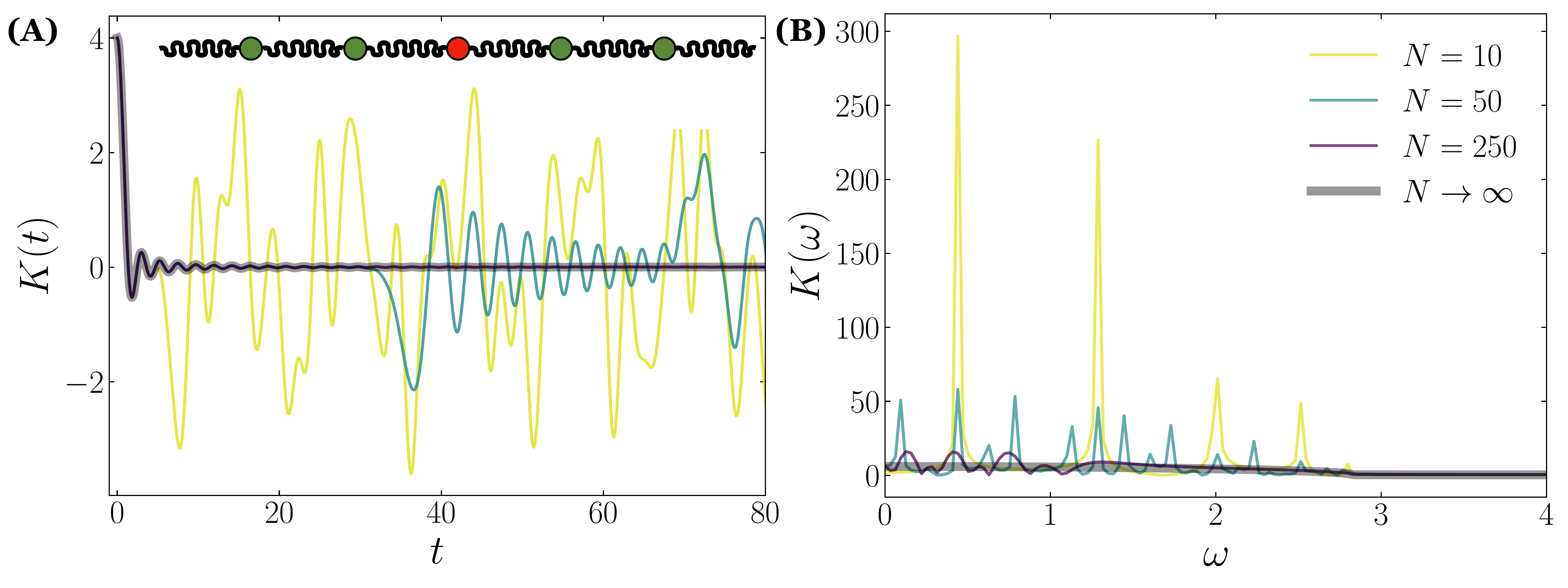}
           \caption{Comparison of memory kernel (A) and power spectra (B) for single site fluctuations of a 1D harmonic chains of various lengths with periodic boundary.}
       \label{figure_s5}
    \end{figure}
    
    In Figure \ref{figure_s5} we illustrate the size dependence of the memory kernel for a site in a 1D chain. Like the 3D lattices presented in the main text, there is a frequency shift as we move to increase the size of the chain. However, the power spectra is not bimodal, but rather a continuous sum of many modes which decrease in amplitude as we approach the chain's Debye frequency $2 \omega$. Furthermore, the memory kernel also does not decay exponentially, but rather as $\frac{1}{t}$, perhaps a consequence of the well-known ergodicity breaking in such systems.  
    \clearpage
    
\section{Results for other metals and surfaces}
    \label{sec:other_metals}
\begin{figure}[htb]  
       \centering
       \includegraphics*[width=6.0in]{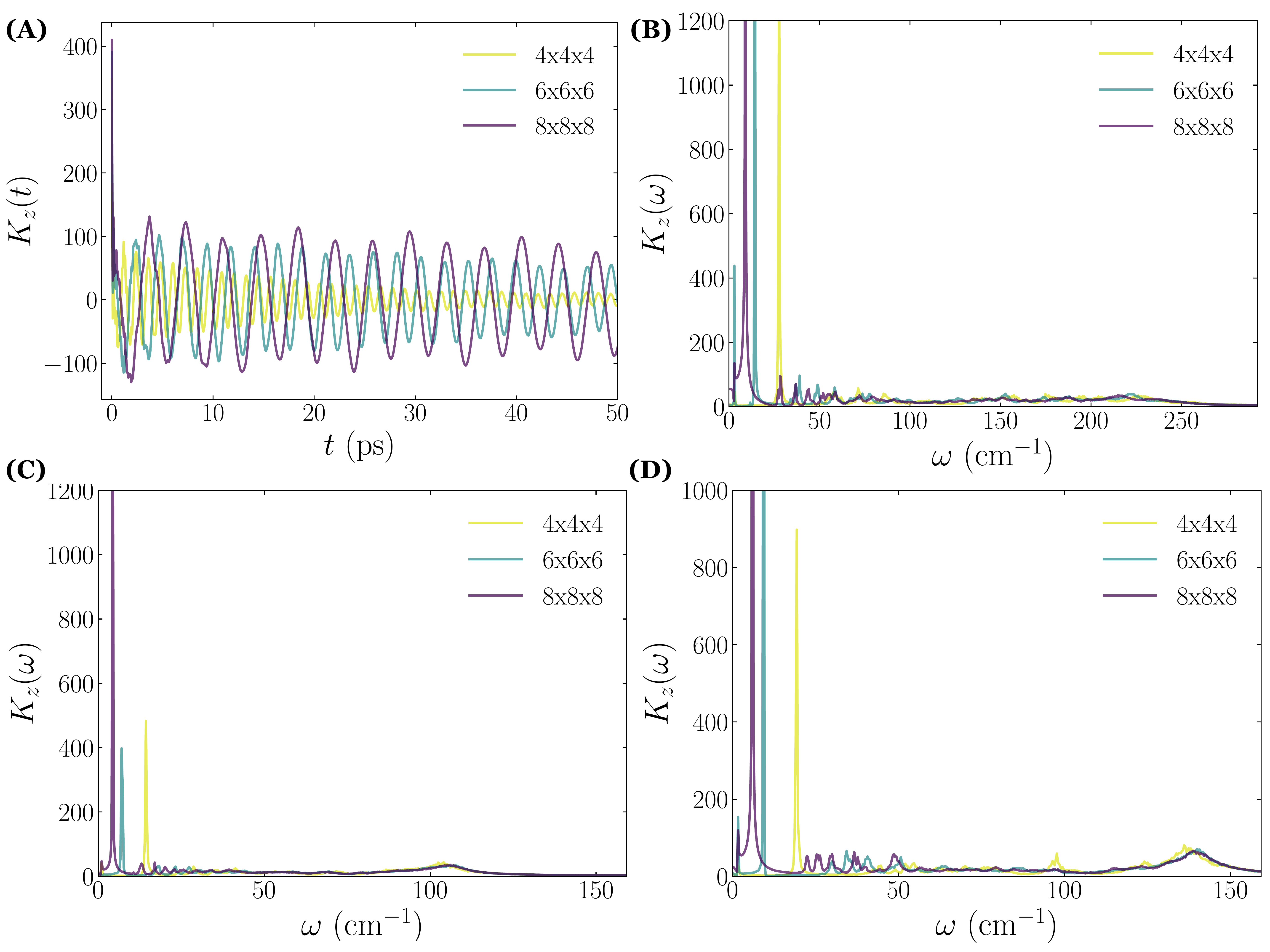}
       \caption{(A) Memory kernel for surface sites of Cu(111) lattices of various sizes simulated with EMT. (B) Associated power spectra of Cu(111). (C) Power spectra for Au(111) lattices simulated with EMT. (D) Power spectra for Pt(100) lattices simulated with EMT. }
       \label{figure_s6}
    \end{figure}

    In Figure \ref{figure_s6} we illustrate the power spectra of metal surfaces other than Pt(111). The power spectra of Au(111) and Pt(110) (Figure \ref{figure_s6}C and Figure \ref{figure_s6}D) clearly have the same bimodal behavior as Pt(111). The power spectra of Cu(111) (Figure \ref{figure_s6}B) appears to be missing the Debye mode, however the real-time memory kernel (Figure \ref{figure_s6}A) has the same characteristic fast decay followed by coherent oscillations which decay much slower. Comparing Figure \ref{figure_s6}A and Figure \ref{figure_s6}B suggests that surface sites still do couple to modes near the Debye frequency in Cu(111), however such modes are of such a high frequency and dissipate energy so quickly that they are either overdamped, or not properly resolved due to numerical errors when computing the correlation functions. This suggestion is corroborated by the fact that the experimental Debye frequency of Cu is nearly two times greater than that of Au, and roughly 50\% greater than that of Pt. 
    
    \FloatBarrier
    \clearpage

\section{Argon scattering simulations}

    \begin{figure}[htb]  
       \centering
       \includegraphics*[width=6.0in]{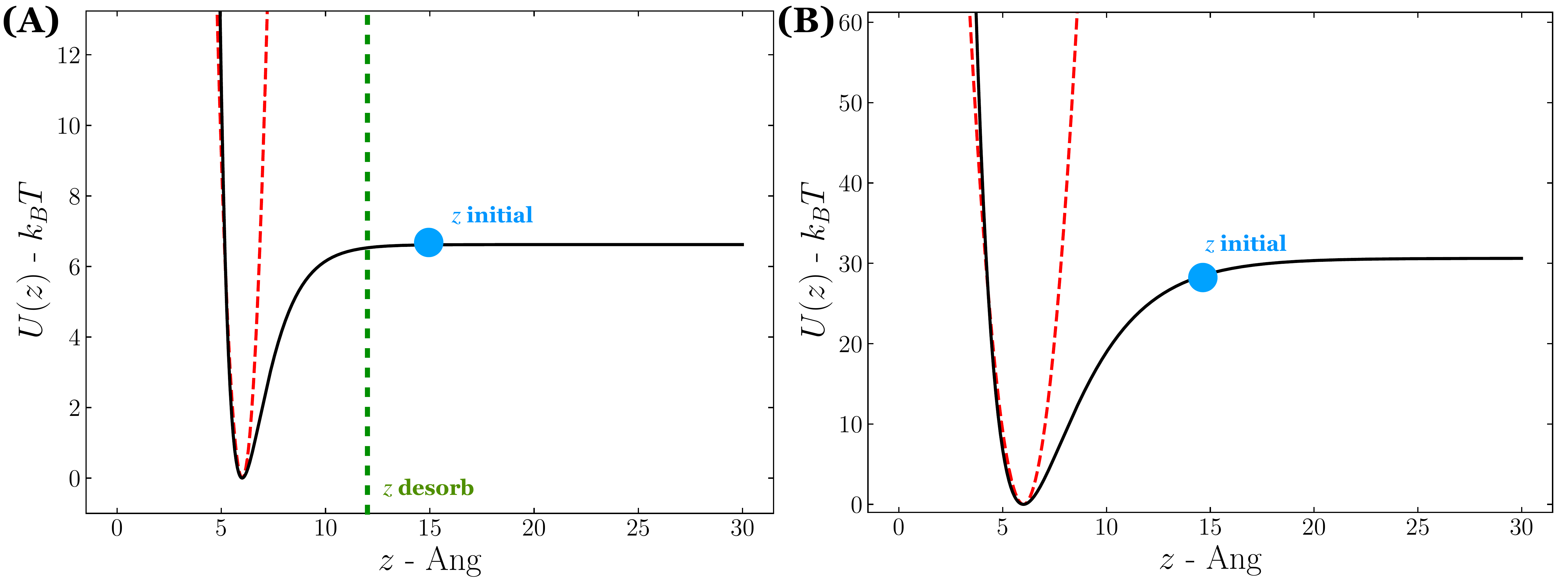}
       \caption{Morse potentials used in main text. The red lines are the harmonic fits to the potential in the well. Blue dots are set at the initial position for scattering simulation. The dark green line sets the target value for surface desorption simulations}
       \label{figure_s7}
    \end{figure}
    
    Figure~\ref{figure_s7} illustrates the Morse potentials used in the main text to describe the interaction of Argon and a platinum surface. In Figure~\ref{figure_s7}B we increase the depth of the well $D$, keeping the frequency $\omega_z$ the same. The exact parameters of these potentials are shown in Table S2.
    
    \begin{table}[htb]
        \caption{Morse potential parameters between Argon on Pt(111) surface used in main text. Taken from Ref. 43}
        \label{tab:morse}
        \begin{tabularx}{\textwidth}{XXXX}
            \toprule
             & D (eV) & a ($\mathrm{Ang^{-1}}$) & $\omega_z$ ($\mathrm{cm^{-1}}$)  \\ \midrule
            Ar & 6.62 & 44 & 0.83 \\
            Ar - deep well & 30.62 & 44 & 0.39\\ \bottomrule
        \end{tabularx}
    \end{table}    

    In Figure S7 we present the power spectra of the GLE models used to conduct the surface scattering and desorption simulations in the main text. The 5 term fit includes some of the more minor peaks of the power spectra not included in the 2 term fit. The macroscopic limit only includes the Debye peak.

    \begin{figure}[htb]  
       \centering
       \includegraphics*[width=6.0in]{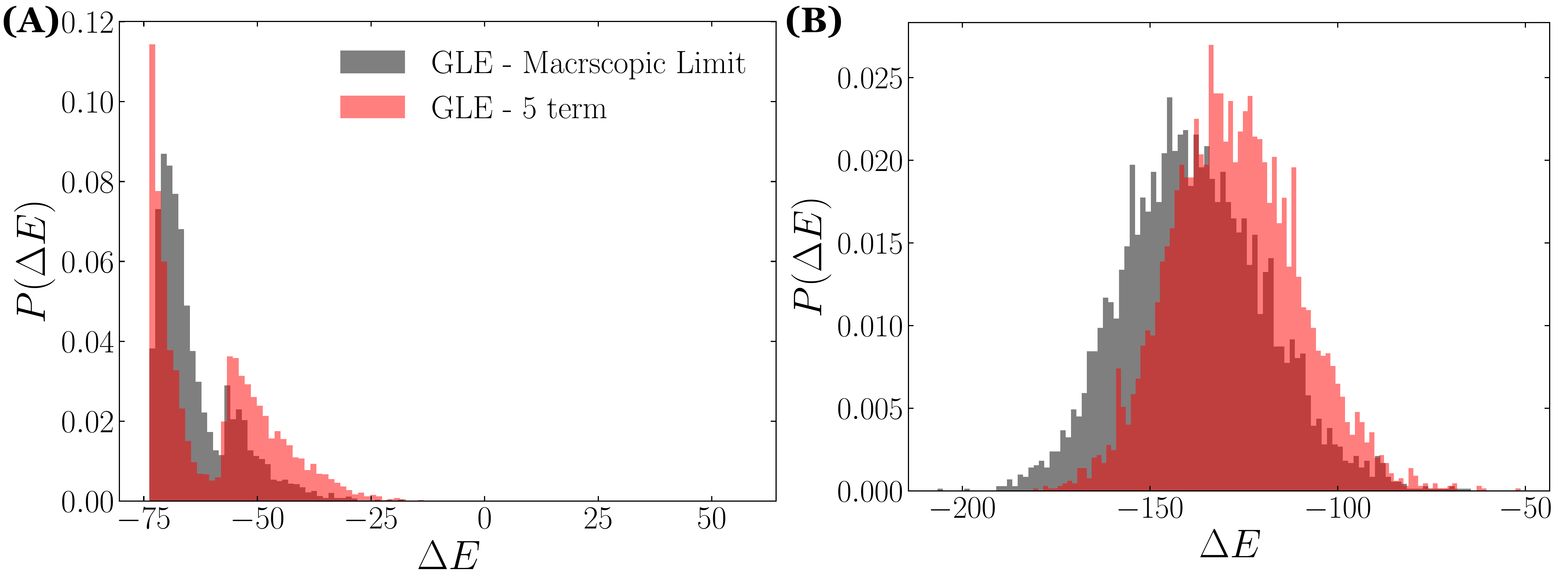}
       \caption{ Histograms for energy dissipated during scattering using $D=30.62$ eV Morse potential. (A) Using trajectories with $\mathrm{KE}(t=0)/D = 0.75$. (B) Using trajectories with incident KE to well-depth ratio of $\mathrm{KE}(t=0)/D = 2.5$.}
       \label{figure_s8}
    \end{figure}
    In Figure S8 we present histograms of the energy lost from the adsorbate to the lattice during the surface scattering simulations using the $D=30.62$ eV potential at two different values of the incident velocity. At low incident velocities, the distribution is bimodal and asymmetric, however as we increase the incident velocity, the distribution becomes increasingly Gaussian. The bimodality at low incident velocities is a consequence of some of the trajectories being trapped, and other escaping. Trapped trajectories interact for longer with the lattice and therefore dissipate more energy. 

    Interestingly, when $\mathrm{KE}(t=0)/D = 2.5$ all the trajectories escape, however, by analyzing the histograms in Figure S8B we see can still see a signature of the finite-size effects discussed in the main text. The 5-term GLE model dissipates less energy than the macroscopic limit GLE model due to coupling to acoustic modes. 
    \clearpage
    
\section{Argon desorption simulations}
    
    In barrier crossing simulations we begin an ensemble of 2000 trajectories situated at $z=z_0$, the minimum of the Morse potential, with a velocity sampled from a thermal normal distribution at 300 Kelvin. We then compute the desorption rate constant as the mean flux out of the "adsorbed state" over population in this state,
    \begin{equation}
        k_d = - \left\langle \frac{ d P_a }{dt} \right\rangle / P_a,
    \end{equation}
    where $k_d$ is the desorption rate constant, $P_a$ is the population in the well or adsorbed state, and $\frac{dP_a}{dt}$ is the change or flux of this population. A trajectory is set to have "desorbed"  when it reaches roughly 18 times the Morse well-width parameter $\frac{1}{\alpha}$, which corresponds to 12 Angstroms, for the first time. We verified that variations in this cutoff from 6-20 Angstroms play no role in the final results. Once a trajectory has reached the desorbed state, it is ended and another trajectory is started at $z_0$. This keeps the population in the well roughly constant. 
    
    \begin{figure*}[h]  
       \centering
       \includegraphics*[width=3.25in]{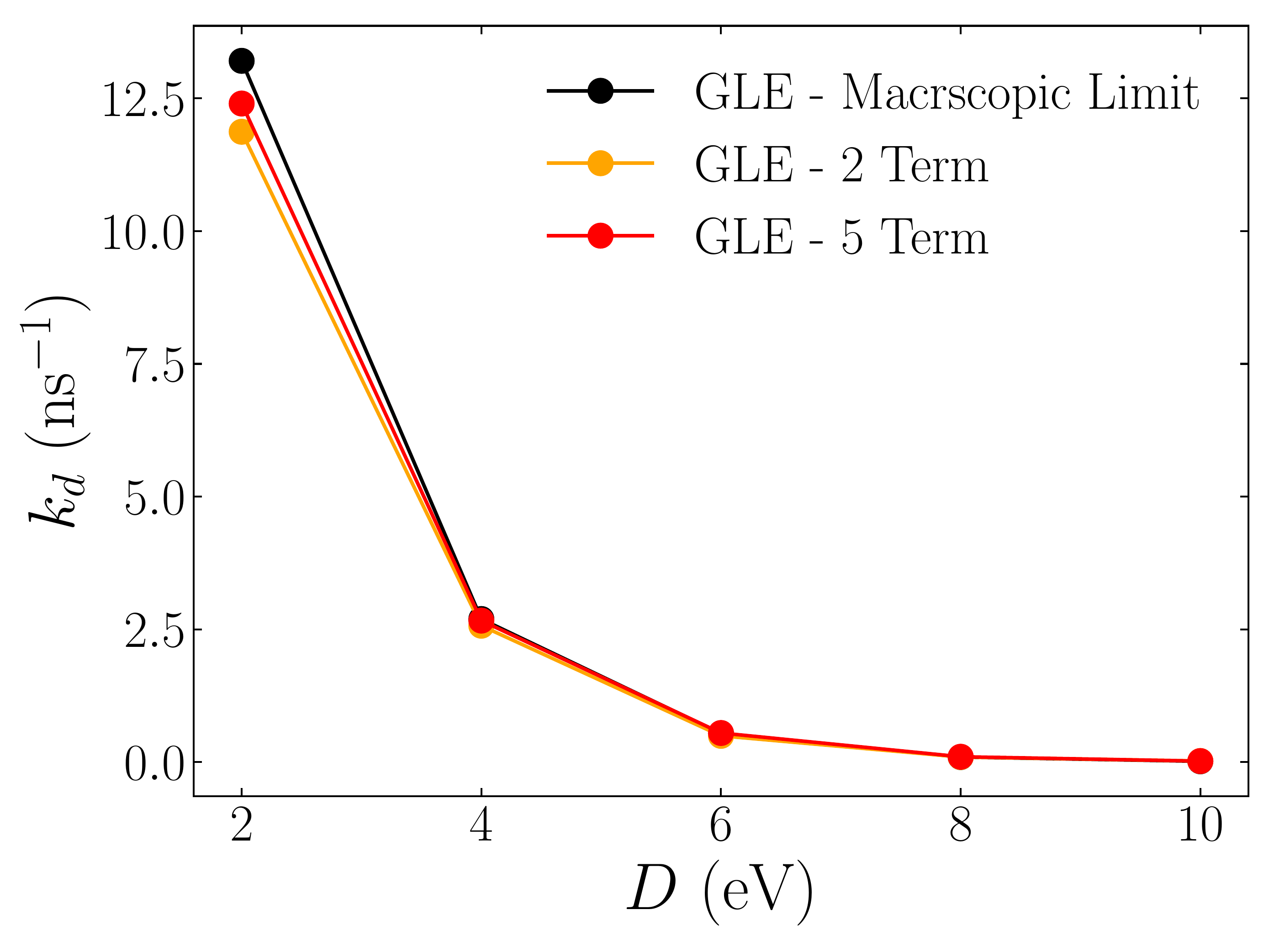}
       \caption{ Desorption rate constants $k_d$ as a function of well-depth $D$. }
       \label{figure6}
    \end{figure*}
    In contrast to our simulations of surface scattering, our simulations of barrier crossing exhibit only a very minor difference between the results for the nanoscale GLE models and the extrapolated macroscopic limit (Figure 5.). One possible explanation is the difference in initial conditions between scattering and barrier crossing simulations. In the scattering simulations the adsorbate begins in a non-equilibrium state and we observe it's relaxation, meanwhile in barrier crossing simulations the adsorbate begins near the equilibrium state and the desorption rate constant arises from fluctuations out of that state.